# Multiscale modeling strategy and general theory of non-equilibrium plasma assisted ignition and combustion

Suo Yang[1], Wenting Sun[2] and Vigor Yang[3]

*School of Aerospace Engineering*

*Georgia Institute of Technology, Atlanta, GA 30332*

## ABSTRACT

A self-consistent 1D theoretical framework for plasma assisted ignition and combustion is reviewed. In this framework, a "frozen electric field" modeling approach is applied to take advantage of the quasi-periodic behaviors of the electrical characteristics to avoid the re-calculation of electric field for each pulse. The correlated dynamic adaptive chemistry (CO-DAC) method is employed to accelerate the calculation of large and stiff chemical mechanisms. The time-step is updated dynamically during the simulation through a three-stage multi-timescale modeling strategy, which takes advantage of the large separation of timescales in nanosecond pulsed plasma discharges. A general theory of plasma assisted ignition and combustion is then proposed. Nanosecond pulsed plasma discharges for ignition and combustion can be divided into four stages. Stage I is the discharge pulse, with timescales of O(1-10 ns). In this stage, most input energy is coupled into electron impact excitation and dissociation reactions to generate charged/excited species and radicals. Stage II is the afterglow during the gap between two adjacent pulses, with timescales of O(100 ns). In this stage, quenching of excited species not only further dissociates $O_2$ and fuel molecules, but also provides fast gas heating. Stage III is the remaining gap between pulses, with timescales of O(1-100 μs). The radicals generated during Stages I and II significantly enhance the exothermic reactions in this stage. Stage IV is the accumulative effects of multiple pulses, with timescales of O(1 ms – 1 sec), which include preheated gas temperatures and a large pool of radicals and fuel fragments to trigger ignition. For plasma assisted flames, plasma significantly enhances the radical generation and gas heating in the pre-heat zone, which could trigger upstream auto-ignition.

[1] Ph.D. Candidate and Graduate Research Assistant, *syang305@gatech.edu*
[2] Assistant Professor
[3] William R. T. Oakes Professor and Chair

**Nomenclature**

$C_{p,k}$ = *specific heat at constant pressure of the k-th species* ($J \cdot kg^{-1} \cdot K^{-1}$)

$D_k$ = *effective diffusion coefficient of the k-th species* ($cm^2 \cdot s^{-1}$)

$D_\epsilon$ = *electron energy diffusion coefficient* ($cm \cdot V^{-1} \cdot s^{-1}$)

$Dec_k$ = *decay rate of the k-th species* ($kg \cdot cm^{-3} \cdot s^{-1}$)

$\boldsymbol{E}$ = *electric field* ($V \cdot cm^{-1}$)

$E$ = *total energy* ($J \cdot kg^{-1}$)

$\boldsymbol{E}/N$ = *reduced electric field (Td)*

$F_i^{EHD}$ = *electro-hydrodynamic force per unit volume* ($kg \cdot cm^{-2} \cdot s^{-1}$)

$G(t)$ = *non-dimensional heat transfer parameter*

$h$ = *gas mixture enthalpy* ($J \cdot kg^{-1}$)

$h_k^0$ = *enthalpy of formation of the k-th species at* $T_{ref}$ ($J \cdot kg^{-1}$)

$\boldsymbol{J}_e$ = *electron density flux* ($cm^{-2} \cdot s^{-1}$)

$\boldsymbol{J}_{e,s}$ = *wall boundary flux of electrons (cm$^{-2}$·s$^{-1}$)*

$\boldsymbol{J}_k$ = *flux of the $k$-th species (cm$^{-2}$·s$^{-1}$)*

$\boldsymbol{J}_\epsilon$ = *flux of electron energy (eV·cm$^{-2}$·s$^{-1}$)*

$\boldsymbol{J}_{\epsilon,s}$ = *wall boundary flux of electron energy (eV·cm$^{-2}$·s$^{-1}$)*

$\boldsymbol{J}_{+,-}$ = *net positive and negative charge fluxes, respectively (cm$^{-2}$·s$^{-1}$)*

$\boldsymbol{J}_{+,s}$ = *wall boundary flux of positive ions (cm$^{-2}$·s$^{-1}$)*

$\boldsymbol{J}_{-,s}$ = *wall boundary flux of negative ions (cm$^{-2}$·s$^{-1}$)*

$k_d$ = *thermal conductivity of quartz (W·m$^{-1}$·K$^{-1}$)*

$k_{gw}$ = *thermal conductivity of the gas mixture at temperature $T_{gw}$ (W·m$^{-1}$·K$^{-1}$)*

$L$ = *gap length (cm)*

$l_d$ = *thickness of the dielectric layer (cm)*

$M$ = *average molecular weight of air (kg)*

$m_e$ = *electron mass (kg)*

$N$ = *number density of neutral particles (cm$^{-3}$)*

$\boldsymbol{n}_s$ = *outward unit normal vector*

$n_e$ = *electron number density (cm$^{-3}$)*

$n_k$ = *number density of the $k$-th species (cm$^{-3}$)*

$n_\epsilon$ = *electron energy density (eV·cm$^{-3}$)*

$n_{+,-}$ = *sum of number densities of positive and negative ions, respectively (cm$^{-3}$)*

$p$ = *pressure (kg·cm$^{-1}$·s$^{-1}$)*

$\dot{Q}_\epsilon$ = *production rate of electron energy density (eV·cm$^{-3}$·s$^{-1}$)*

$\dot{Q}^{JH}$ = *energy release rate from Joule heating (kg·cm$^{-1}$·s$^{-3}$)*

$Q_{pulse}$ = *input energy per pulse (mJ)*

$q_i$ = *energy flux from heat conduction and diffusion (kg·s$^{-3}$)*

$q_k$ = *charge number of the k-th species (-1 for negative ions and electrons, +1 for positive ions, and 0 for neutral species)*

$r_i$ = *reaction rate of the i-th electron impact reaction (cm$^{-3}$·s$^{-3}$)*

$S_L$ = *laminar flame speed (cm·s$^{-1}$)*

$S_{ign}$ = *ignition kernel propagation speed (cm·s$^{-1}$)*

$T$ = *gas temperature (K)*

$T_{amb}$ = *ambient temperature (K)*

$T_b$ = *boundary temperature (K)*

$T_e$ = *electron temperature (K)*

$T_{gw}$ = *gas temperature at a distance Δx from the solid wall (K)*

$T_{ref}$ = *reference temperature (K)*

$T_{se}$ = *temperature of secondary electrons ejected from the electrode surface (eV)*

$t_{pulse}$ = *duration of one pulse (ns)*

$\boldsymbol{u}$ = *convective velocity vector of the gas mixture (cm·s$^{-1}$)*

$u_i$, $u_j$ = *flow velocity components in* $i$*-th and* $j$*-th directions, respectively (cm·s$^{-1}$)*

$V_{app}$ = *applied voltage (V)*

$V_{gap}$ = *gap voltage (V)*

$V_{peak}$ = *peak value of applied voltage (kV)*

$\gamma$ = *secondary electron emission coefficient for ions colliding with electrode surface*

$\Delta E_i$ = *heat of the i-th electron impact reaction (eV)*

$\Delta n_{e,max}$ = *maximum relative difference in the electron number density at the end of two adjacent discharge pulses*

$\epsilon$ = *electric permittivity (F·cm⁻¹)*

$\epsilon_d$ = *dielectric constant*

$\epsilon_e$ = *electron energy (eV)*

$\kappa$ = *dynamic viscosity* ($kg \cdot cm^{-1} \cdot s^{-1}$)

$\lambda$ = *thermal conductivity of the gas mixture* ($W \cdot cm^{2} \cdot K^{-1}$)

$\mu_k$ = *mobility of the k-th species in the electric field* ($cm^{2} \cdot V^{-1} \cdot s^{-1}$)

$\mu_\epsilon$ = *electron energy mobility* ($cm^{2} \cdot V^{-1} \cdot s^{-1}$)

$\nu_{el}$ = *elastic collision frequency of electrons* ($s^{-1}$)

$\rho$ = *density of the plasma mixture* ($kg \cdot cm^{-3}$)

$\tau_{ij}$ = *viscous shear stress tensor* ($kg \cdot cm^{-1} \cdot s^{-2}$)

$\tau_{plasma}$ = *ignition delay after plasma pulse burst (s)*

$\tau_{self}$ = *auto-ignition delay without plasma (s)*

$\phi$ = *electric potential (V)*

$\dot{\omega}_k$ = *production term of the $k$-th species* ($cm^{-3} \cdot s^{-1}$)

## 1. Introduction

Over the past decade, non-equilibrium plasma has been the subject of significant attention, due to its great potential to enhance ignition and combustion in internal combustion engines, gas turbines, scramjet engines, and pulsed detonation engines [1, 2]. Past studies have shown that non-equilibrium plasma can shorten ignition delays [3, 4], extend the flammability limits to allow ultra-lean combustion for emission reduction [5], increase flame propagation speed [6, 7], and improve flame stabilization [8]. While the phenomena observed in these investigations are very promising, however, the underlying physio-chemical processes for such enhancement are still not well understood. In order to understand these underlying processes, a combination of experimental and numerical investigations is necessary.

The existing experimental studies can be divided into two categories. The first category reproduces the complicated reacting flow conditions in engines. For example, Zhang *et al.* [9, 10] investigated the interaction between a plasma jet and turbulent flows; Kim *et al.* [11] investigated the interaction between plasma and a fuel jet in a cross flow; Starikovskaia *et al.* [12] and Leonov *et al.* [13, 14] investigated the interaction between plasma and supersonic reacting flow. Due to the complicated thermal, kinetic, and transport coupling contexts of these configurations, however, this line of research can offer only limited elucidation of specific plasma enhancement effects. To tackle individual issues, experiments in well-studied simple configurations have been used to simplify or even eliminate hydrodynamic effects and isolate plasma enhancement from other effects. For example, Uddi *et al.* [15] conducted two-photon absorption laser induced fluorescence (TALIF) measurements of atomic oxygen in fuel/air mixtures subjected to nanosecond pulsed discharges in a rectangular quartz reactor. Yin *et al.* [16] studied the ignition of mildly preheated $H_2$/air mixtures under nanosecond pulsed discharges in a quartz flow reactor. Lefkowitz *et al.* [17, 18] conducted *in situ* measurements of nanosecond pulsed plasma activated $C_2H_4$/Ar pyrolysis and oxidation of $C_2H_4$/$O_2$/Ar mixtures in a flow reactor.

Although this second group of experimental studies can provide more insights into the underlying physio-chemical processes of plasma enhancement, the number of measurable quantities is very limited and many conclusions can only be indirectly inferred. For this reason, high-fidelity modeling and simulation of plasma assisted ignition and combustion is vital. Comprehensive numerical models of plasma have been developed over the last several decades, and the field is relatively mature. For example, Ventzek *et al.* [19] developed a high dimensional plasma model, and Shigeta [20] reviewed a class of models for plasma-turbulence interaction. In contrast, the modeling of plasma assisted ignition and combustion is very limited, primarily because the range of scales involved makes the computation extremely time-consuming. In addition, most of these studies [21-23] are zero-dimensional (0D), due to the extreme complexity of plasma-combustion interaction. However, in 0D models, plasma discharges are assumed to be uniform over the entire domain during each pulse, and the influence of cathode sheath formation is neglected. Furthermore, because the energy input channels are difficult to model in 0D models, the reduced electric field (E/N) and electron density are pre-specified such that the coupled energy can match the experimentally measured values. To address these issues, our group has developed a self-consistent, one-dimensional (1D), multiscale numerical model [24-26] for plasma assisted ignition and combustion. This model resolves the transient electric field during each nanosecond discharge pulse, and calculates the cumulative effects of multiple pulses on fuel dissociation/pyrolysis, oxidation, ignition, and combustion. The model has been extensively validated in three typical configurations for a wide range of fuel types and operating conditions [27-31]. In the first half of this paper, we will summarize the different modeling strategies we have developed over the past decade.

There have been several topic reviews for the field of plasma assisted ignition and combustion published in recent years. Starikovskaia [12] and Starikovskii *et al.* [32] summarized the discharge types, and reviewed the oxidation of different hydrocarbons by nanosecond discharge and the control of their ignition and combustion. Starikovskiy and Aleksandrov [1] reviewed the applications of plasma assisted combustion and its physics. Sun and Ju [33-35] reviewed the progress in chemical kinetics studies and advanced diagnostics of plasma assisted low temperature combustion. The present paper focuses on numerical modeling and a general multi-timescale theory of plasma assisted ignition and combustion.

The structure of this paper is as follows. Section 2 reviews our self-consistent theoretical framework for plasma assisted ignition and combustion. Section 3 summarizes the different modeling strategies. Finally, Section 4 develops a general theory for nanosecond plasma discharges, plasma assisted ignition, and plasma assisted premixed and non-premixed flames.

## 2. Theoretical Framework

This section covers the physical configurations under consideration, the governing equations and their boundary conditions, the necessity for at least one-dimensional modeling, and plasma-combustion chemistry.

### 2.1 Physical Configurations

Three configurations are considered in this study to cover the ignition, premixed flame, and non-premixed flame, under non-equilibrium plasma discharges.

The first configuration [24, 27-29], shown in Fig. 1(a), is designed to simulate plasma discharges of fuel/oxidizer/diluent mixtures for low temperature fuel oxidation and ignition. It is a plane-to-plane geometry to match nanosecond dielectric barrier discharge (NS DBD) reactor experiments [16, 18, 29, 36]. Two copper electrodes covered by thin dielectric layers are placed 1-1.5 cm apart, with the gap between them filled with gas mixtures. A high voltage power supply is connected to the right electrode, while the left electrode is grounded.

The second configuration [30], shown in Fig. 1(b), simulates a plasma assisted premixed flame. It is a steady laminar flat premixed flame directly coupled with plasma discharge to match with the McKenna burner experiment [37]. The entire flame (preheat zone, reaction zone, and post-flame zone) is encapsulated in the plasma discharge. The burner exit serves as the ground electrode, and the high-voltage electrode, which is a perforated metal plate, is located 4 cm above the burner exit.

The third configuration [31, 38], shown in Fig. 1(c), simulates a plasma assisted non-premixed flame in counter-flow experiment [39]. Two wire-mesh electrodes covered with thin alumina as dielectric layer (1.5 mm in thickness) are placed 1 cm apart in a tube filled with oxidizer. In this configuration, the discharge region is separated from the oxidizer burner exit. Thus, the residence time is large enough for gas temperature and species concentration to become uniform at the cross section of the oxidizer burner exit, to match the 1D assumption in this study.

(a)

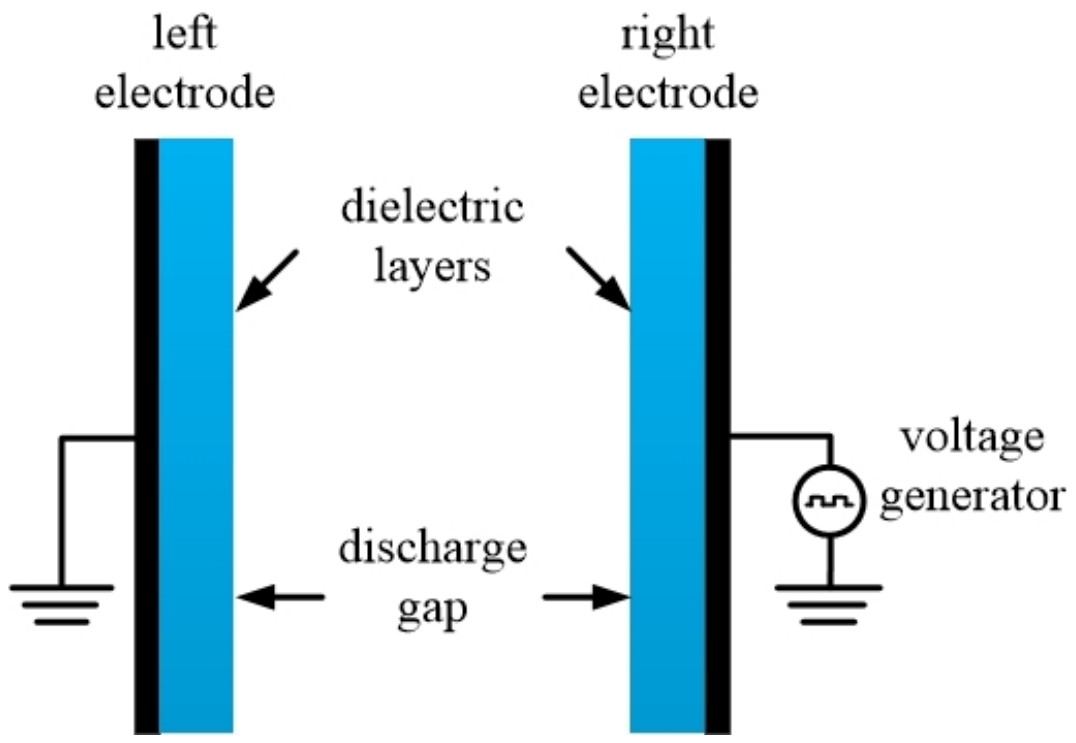

(b)

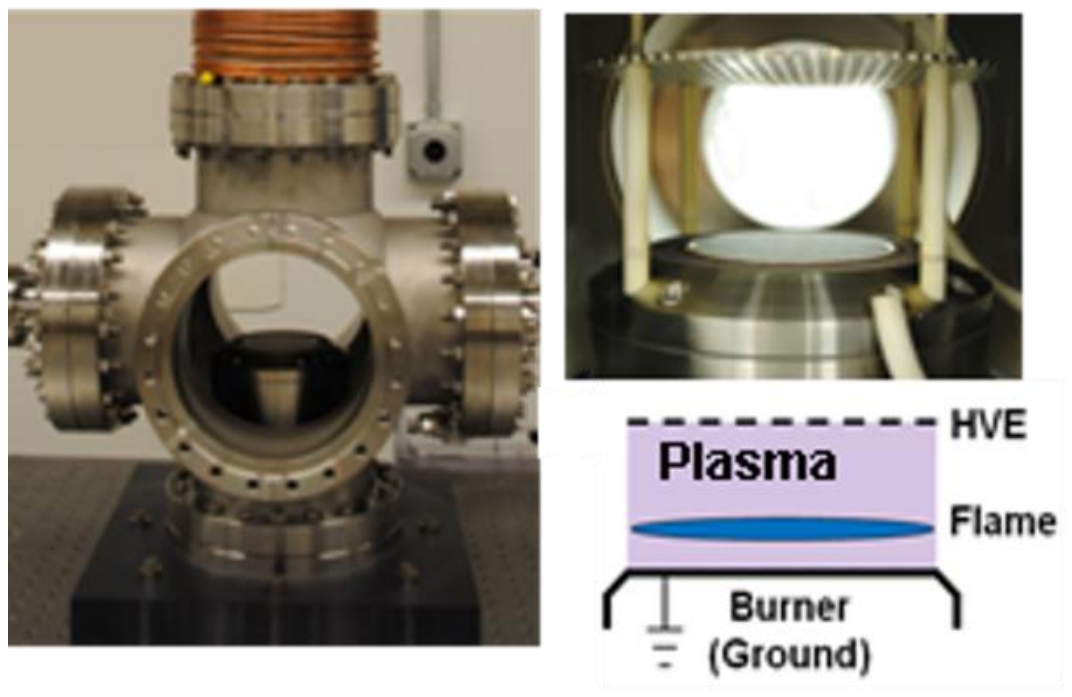

(c)

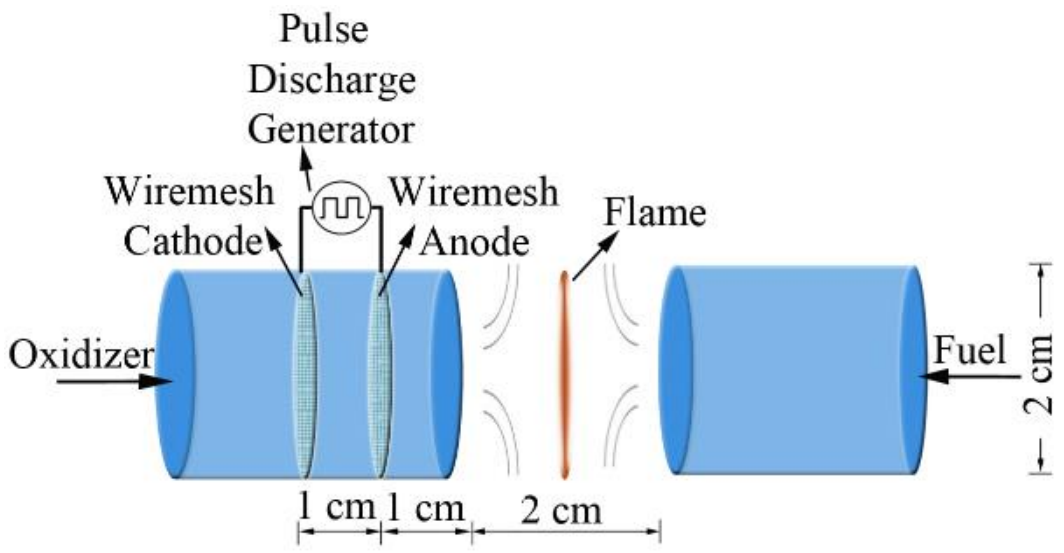

**Figure 1.** (a) Schematic of plane-to-plane simulation configuration [24, 27-29]. (b) Schematic of the "direct coupling" flat premixed flame configuration [30]. (c) Schematic of counter-flow non-premixed flame configuration [31].

### 2.2 Governing Equations

In this study, we use a two-temperature non-equilibrium plasma model, in which ions and neutral species are in thermal equilibrium at the gas temperature $T$ [40], and electrons are in thermal non-equilibrium with the electron temperature $T_e$ approximated by the "local electron mean energy method" [41]. The governing equations include the Poisson equation for electric potential, the electron energy equation, and detailed finite rate transport equations for charged and neutral species as Eqs. (1)-(3):

$$\boldsymbol{\nabla} \cdot (\epsilon \boldsymbol{\nabla} \phi) = -e(n_+ - n_- - n_e) \tag{1}$$

$$\frac{\partial n_\epsilon}{\partial t} + \boldsymbol{\nabla} \cdot \boldsymbol{J}_\epsilon = \dot{\mathrm{Q}}_\epsilon \tag{2}$$

$$\frac{\partial n_k}{\partial t} + \boldsymbol{\nabla} \cdot \boldsymbol{J}_k = \dot{\omega}_k \tag{3}$$

Electric field ($\boldsymbol{E} = -\nabla\phi$) and energy input is thus calculated, rather than pre-specified as in the 0D models [18]. The transport of energy and species is calculated using the drift (mobility)-diffusion model. The electron energy density is defined as $n_\epsilon = n_e \epsilon_e$. The flux term $\boldsymbol{J}_\epsilon$ is defined as $\boldsymbol{J}_\epsilon = -\mu_\epsilon n_\epsilon \boldsymbol{E} - \nabla(D_\epsilon n_\epsilon) + n_\epsilon \boldsymbol{u}$ . The source term $\dot{Q}_\epsilon$ is defined as $\dot{Q}_\epsilon = -\left(\frac{3k_B m_e}{eM}\right) n_e \nu_{el}(T_e - T) - \sum \Delta E_i r_i - \sum \boldsymbol{J_e} \cdot \boldsymbol{E}$. Here, the first term "$-\left(\frac{3k_B m_e}{eM}\right) n_e \nu_{el}(T_e - T)$" represents the electron energy loss from elastic collisions with gas molecules. The second term "$-\sum \Delta E_i r_i$" represents the energy loss from electron impact reactions. The third term "$-\sum \boldsymbol{J_e} \cdot \boldsymbol{E}$" represents the energy gain from the acceleration in the applied electric field. For species transport equations, the flux term $\boldsymbol{J}_k$ is defined as $\boldsymbol{J}_k = q_k \mu_k n_k \boldsymbol{E} - \nabla(D_k n_k) + n_k \boldsymbol{u}$. The electron transport and reaction-rate coefficients are fitted as functions of electron energy $\epsilon_e$ calculated by solving the electron Boltzmann equation with two-term expansion using the BOLSIG software [42], and renewed by interpolation at every time step and every spatial grid-point.

The gas flow is governed by the conservation equations of mass, momentum and total energy, as Eqs. (4)-(6), respectively:

$$\frac{\partial \rho}{\partial t} + \frac{\partial \rho u_i}{\partial x_i} = 0 \tag{4}$$

$$\frac{\partial \rho u_i}{\partial t} + \frac{\partial (\rho u_i u_j)}{\partial x_j} = -\frac{\partial p}{\partial x_i} + \frac{\partial \tau_{ij}}{\partial x_j} + F_i^{EHD} \tag{5}$$

$$\frac{\partial \rho E}{\partial t} + \frac{\partial [(\rho E + p) u_i]}{\partial x_i} = -\frac{\partial q_i}{\partial x_i} + \frac{\partial (u_i \tau_{ij})}{\partial x_j} + \dot{Q}^{JH} \tag{6}$$

Following classical fluid mechanics, the total energy $\rho E$ is defined as $\rho E = \rho h - p + \frac{\rho(u_i u_j)}{2}$, where the mixture enthalpy is defined as $h = \sum_k Y_k \left\{h_k^0(T_{ref}) + \int_{T_{ref}}^{T} C_{p,k}(T')dT'\right\}$. The stress tensor $\tau_{ij}$ is defined as $\tau_{ij} = \kappa\left(\frac{\partial u_i}{\partial x_j} + \frac{\partial u_j}{\partial x_i}\right)$, and the energy flux $q_i$ from heat conduction and species diffusion is defined as $q_i = -\lambda\left(\frac{\partial T}{\partial x_j}\right) + \rho \sum_k h_k Y_k D_k \left(\frac{\partial Y_k}{\partial x_j}\right)$. Two special terms are introduced for plasma. The electro-hydrodynamic force term $F_i^{EHD}$ is defined as $F_i^{EHD} = eE_i(n_+ - n_- - n_e)$. The Joule heating rate $\dot{Q}^{JH}$ is defined as $\dot{Q}^{JH} = e\boldsymbol{E} \cdot \sum(\boldsymbol{J}_+ - \boldsymbol{J}_- - \boldsymbol{J}_e)$. Unlike equilibrium plasma, which transfers electrical energy only into sensible enthalpy, non-equilibrium plasma transfers electrical energy into the total energy of the gas mixture, and this triggers both gas temperature rise and chemical reactions. Heat release from chemical reactions is implicitly included in the unsteady term of the total energy equation, in the form of chemical energy being converted into sensible enthalpy.

### 2.3 Boundary Conditions

Electric potential $\phi$ is set to zero at the left boundary, and to the gap voltage $V_{gap}$ at the right boundary. Gap voltage $V_{gap}$ is calculated from the applied voltage $V_{app}$ by solving the following equation [43]:

$$\frac{dV_{app}}{dt} = \left(1 + \frac{2l_d}{\epsilon_d L}\right)\frac{dV_{gap}}{dt} - \frac{2l_d e}{\epsilon_d \epsilon_0 L}\int_0^L [J_+ - J_-]dx \tag{7}$$

In the simulations, as shown in Fig. 2, a Gaussian fit of the experimental high voltage pulse waveform is used as the applied voltage $V_{app}$. It is worth noting that the input pulse energy and ignition characteristics have been found to be highly sensitive to uncertainties in dielectric properties [27].

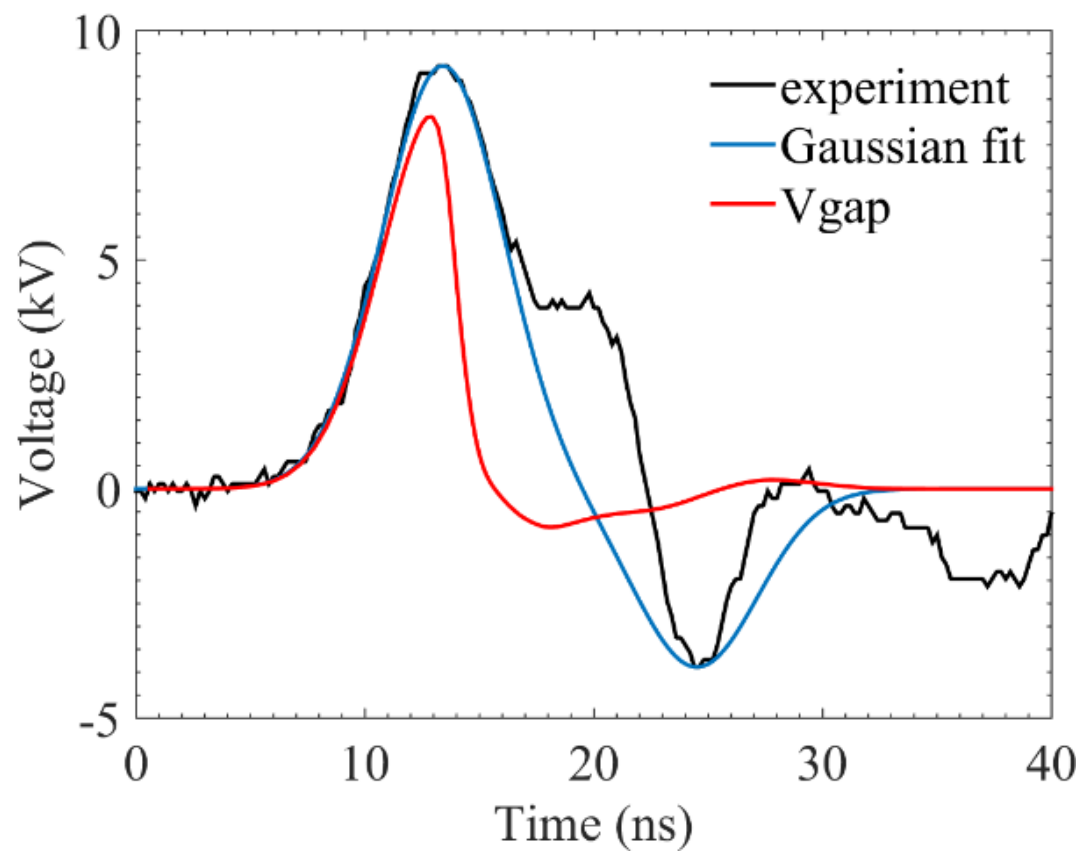

**Figure 2.** Example of Gaussian fit of experimental pulse waveform used in simulation [28].

A zero wall flux is enforced for all neutral species. The wall fluxes for electrons, positive ions, negative ions, and electron energy are specified by Eqs. (8)-(11), respectively [44]:

$$\boldsymbol{J}_{e,s} \cdot \boldsymbol{n}_s = \frac{1}{4} n_e \sqrt{\frac{8k_b T_e}{\pi m_e}} + (a-1)\mu_e n_e \boldsymbol{E} \cdot \boldsymbol{n}_s - a \sum_k \gamma \boldsymbol{J}_{+k,s} \cdot \boldsymbol{n}_s \quad (8)$$

$$\boldsymbol{J}_{+,s} \cdot \boldsymbol{n}_s = \frac{1}{4} n_+ \sqrt{\frac{8k_b T_g}{\pi m_+}} + a\mu_+ n_+ \boldsymbol{E} \cdot \boldsymbol{n}_s \quad (9)$$

$$\boldsymbol{J}_{-,s} \cdot \boldsymbol{n}_s = \frac{1}{4} n_- \sqrt{\frac{8k_b T_g}{\pi m_-}} + (a-1)\mu_- n_- \boldsymbol{E} \cdot \boldsymbol{n}_s \quad (10)$$

$$\boldsymbol{J}_{\epsilon,s} \cdot \boldsymbol{n}_s = \left(\frac{5}{2} k_b T_e\right)\left[\frac{1}{4} n_e \sqrt{\frac{8k_b T_e}{\pi m_e}} + (a-1)\mu_e n_e \boldsymbol{E} \cdot \boldsymbol{n}_s\right] - a\left(\frac{5}{2} k_b T_{se}\right) \sum_k \gamma \boldsymbol{J}_{+k,s} \cdot \boldsymbol{n}_s \quad (11)$$

where the secondary electron emission coefficient $\gamma$ is assumed to be 0.05 [44]. The temperature of the secondary electrons ejected from the electrode surface $T_{se}$ is set to 1 eV [44]. In addition, $a$=1 if $\boldsymbol{E} \cdot \boldsymbol{n}_s < 0$, and $a$=0 otherwise.

Zero wall flux is enforced for mass and momentum equations. The gas temperature boundary condition is set to be the analytic self-similar solutions of transient temperature distribution in a semi-infinite solid with constant heat flux [45]:

$$T_b = \frac{T_{amb} + G(t) \times T_{gw}}{1 + G(t)};\ G(t) = \frac{4k_{gw}\sqrt{\alpha_d t/\pi}}{k_d \Delta x} \tag{12}$$

In practice, this boundary is closer to isothermal than to adiabatic conditions.

### 2.4 Necessity for At Least One-Dimensional (1D) Modeling

The governing equations and boundary conditions described in the previous sections indicate that modeling must be at least 1D to capture cathode sheath formation and to calculate E/N accurately (0D models pre-specify it). As shown in Fig. 3, the sheath is the boundary between the plasma and the electrodes, with large electric field and electron number density gradients. In fact, rapid gas heating through ion Joule effect mostly occurs inside the sheath boundary rather than in the bulk plasma [24]. This rapid heat addition is rapidly dissipated to the wall, because the boundary is closer to isothermal than adiabatic. Conductive heat loss to the walls, together with thermal diffusion into the bulk gas, prevents the overheating of the cathode sheath layer and the formation of ionization instabilities [24]. For this reason, the input energy matching in 0D models often results in over-prediction of the gas temperature of the bulk plasma.

To prove the necessity of 1D modeling, Yang *et al.* [28] directly compared the results from 1D and 0D models of nanosecond pulsed plasma-activated $C_2H_4/O_2/Ar$ mixtures, using the same plasma-combustion chemistry. For all quantities, the 1D model provides significantly better predictions than the 0D model, with respect to experimental measurements. Further analysis indicated that the kinetics pathways of fuel dissociation and oxidation are similar, but different pathways dominate in the two models.

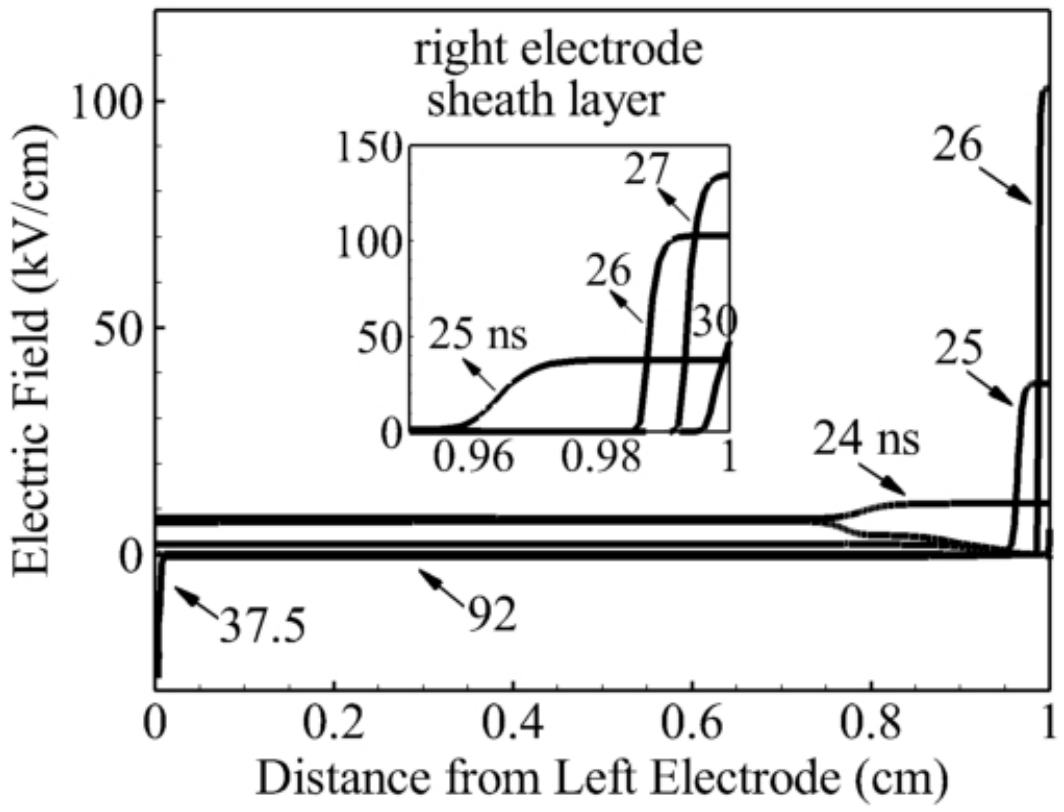

**Figure 3.** Spatial variation in electric field during first voltage pulse. Simulation conducted in air at 60 Torr and 300 K initial pressure and temperature, at 40 kHz repetition rate. [24]

### 2.5 Plasma-Combustion Chemistry

In this study, air or $O_2$/Ar serve as the oxidizer and diluent (background gas), and a wide range of fuels is covered, including hydrogen ($H_2$), ethylene ($C_2H_4$), dimethyl ether (DME: $CH_3OCH_3$), and n-heptane ($nC_7H_{16}$). DME and n-heptane in particular are well-known hydrocarbon fuels with low-temperature (below 700~800 K) chemistry (LTC) behavior. Reaction rate constants of electron impact reactions are calculated at each spatial grid-point and each time step using BOLSIG [42] and modeled as functions of electron energy $\epsilon_e$.

The chemical kinetics models for air plasma [24] and $O_2$/Ar plasma [28] are reduced from a detailed mechanism [15] via sensitivity analysis. The model includes neutral species N, $N_2$, Ar, O, $O_2$, $O_3$, and NO; the charged species $e^-$, $N_2^+$, $N_4^+$, $Ar^+$, $O_2^+$, $O_4^+$, and $O_2^-$; and electronically excited species $N_2(A^3\Sigma)$, $N_2(B^3\Pi)$, $N_2(C^3\Pi)$, $N_2(a'^1\Sigma)$, $Ar^*$, $O_2(a^1\Delta)$, $O_2(b^1\Sigma)$, $O_2(c^1\Sigma)$, $N(^2D)$, and $O(^1D)$. CARS measurements of nitrogen vibrational temperature in air [21] indicated that the vibrational non-equilibrium is insignificant and not likely to affect the plasma chemistry, due to the fast vibrational relaxation within O(10 ns) [46, 47]. Therefore, vibrational excited species are not explicitly included in the chemical kinetics model, and instantaneous vibrational relaxation is assumed. For electron impact reactions, self-consistent sets of electron impact cross-sections are used for $O_2$ [48], $N_2$ [49], and Ar [50]. The rate constants of reactions between $Ar^*$ and $O_2$ were taken from Sun *et al.* [51]. These air plasma chemistry and $O_2$/Ar plasma chemistry models serve as the base component for all the plasma-combustion chemistry mechanisms of fuel/air mixtures and fuel/$O_2$/Ar mixtures.

For $H_2$ as fuel, a chemistry model [27] is reduced from the above-mentioned plasma air chemistry, classical $H_2/O_2$ ignition kinetics [52], and hydrogen plasma reactions [53, 54] through sensitivity analysis. In addition to those species present in the air plasma chemistry, this reduced mechanism also includes neutral species $H_2$, H, OH, $HO_2$, and $H_2O$, and charged species $HN_2^+$, and $H_3O^+$.

To select a better chemistry model for $C_2H_4$, the performance of two models, HP-Mech [18] and USC Mech-II [55], was compared for the prediction of nanosecond pulsed plasma activated $C_2H_4/O_2$/Ar mixtures [28]. The predictions from HP-Mech were found to be better than those of USC Mech-II in terms of comparison with the experimental measurements, mainly because USC Mech-II does not include the high-pressure/low-temperature reaction pathway $OH+C_2H_4 \rightarrow CH_2CH_2OH$. The $C_2H_4$ excitation cross-sections were estimated based on Janev and Reiter's method [56].

For DME and $nC_7H_{16}$ as fuels, two chemical models [29, 57, 58] were developed based on a DME combustion kinetics model [59] and a reduced $nC_7H_{16}$ model [60, 61]. In addition to $H_2$/air plasma chemistry, these mechanisms also include the charged species $CH_3^+$, $CH_3OCH_2^+$, $CH_3OCH_3^+$, $C_7H_{15}^+$, $C_6H_{13}^+$, $C_5H_{11}^+$, and many neutral species. There are, however, no available cross-section data of electron impact reactions for DME and $nC_7H_{16}$. As an approximation, it is assumed that their cross-sections are similar to that of $C_2H_6$ [62], which has a similar molecular structure. A sensitivity analysis was conducted by varying (up to 5 times) the rate constants of electron impact and quenching reactions, and negligible impact on ignition delay predictions was observed [58].

Past studies [63] proved that the dominant reaction pathway for atomic oxygen (O) generation is $e+O_2=e+O+O(^1D)$, which means that almost half of the O produced by plasma is $O(^1D)$. In typical plasma assisted ignition and combustion conditions, the concentrations of hydrocarbons are high. For this reason, the reactions between $O(^1D)$ and hydrocarbons become much faster (approximately 4 times) than the quenching of $O(^1D)$ to O [28], and should be added to the plasma-combustion chemistry models.

## 3. Numerical Modeling Strategies

The transport equations of mass, momentum, energy, and neutral species are spatially discretized using a total variation diminishing (TVD) central scheme [64, 65]. The charged species equations derived from the drift (mobility)-diffusion model are spatially discretized using an exponential Scharfetter-Gummel scheme [66]. After spatial discretization, the original partial differential equation (PDE) system becomes an ordinary differential equation (ODE) system with respect to time. Non-uniform meshes consisting of O(100-1000) grid points are employed to reach convergent solutions, with the mesh clustering toward the two electrodes to resolve the high gradients inside the sheath layers. The model is programmed in parallel on multiple processors via domain decomposition and message passing interface (MPI) techniques.

### 3.1 “Frozen Electric Field” Strategy

The “frozen electric field” strategy takes advantage of the quasi-periodic behaviors of the electrical field to avoid the re-calculation of the electric field for each pulse.

The transport equation of electron number density and the Poisson equation for electric potential are strongly coupled, which results in a severe time-step restriction. To handle this issue without affecting the accuracy, a semi-implicit form of the Poisson equation [67] is solved using the implicit Lower Upper (LU) factorization method. The transport equations of electron number density and electron energy density are time-advanced using an implicit Generalized Minimal Residual (GMRES)-based ODE solver [68].

Under repeated nanosecond discharge, the electrical characteristics (input energy, E/N, $n_\epsilon$, $n_e$, etc.) of the non-equilibrium plasma also present quasi-periodic behaviors, as shown in Figs. 3(a) and 4. Taking advantage of this behavior, the electrical characteristics do not need to be re-calculated every pulse, which significantly reduces computation time. More importantly, the stiffness of the system can be reduced dramatically, and the extremely stringent time step size of O(0.1-1 ps) during each pulse can be relaxed to O(0.1-1 ns). Detailed evaluations [25, 28] show that the electrical characteristics are not exactly periodic, but approach pure periodicity after a certain number of pulses, when the input energy per pulse becomes stable (that is, the field achieves quasi-equilibrium). After the plasma has reached quasi-equilibrium, the accumulated input energy is linearly proportional to the number of pulses. Based on this observation, a “frozen electric field” strategy [25] is implemented.

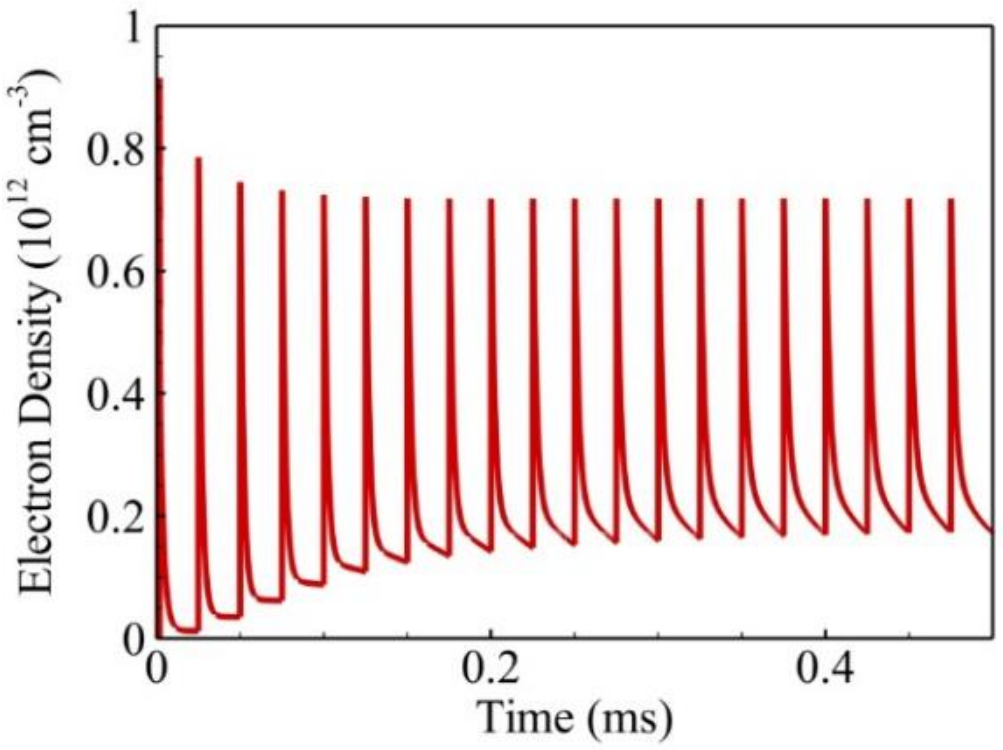

**Figure 4.** Temporal evolution of electron density vs time at the center of the discharge domain. Simulation conducted in $H_2$/air mixture at 114 Torr and 473 K initial pressure and temperature, respectively, at 40 kHz repetition rate. [25]

The "frozen electric field" strategy contains three steps. First, several discharge pulses are simulated based on the full plasma-combustion model, until $\Delta n_{e,max} \leq 5\%$. Second, the temporal-spatial distributions of E/N, $n_\epsilon$, and $n_e$ during the first pulse to reach $\Delta n_{e,max} \leq 5\%$ are saved in a look-up table. Last, in all the following pulses, the Poisson equation for $\phi$ and the transport equations of $n_\epsilon$ and $n_e$ are turned off, with the values of E/N, $n_\epsilon$, and $n_e$ frozen as the values in the table. In most cases, approximately 10 pulses are adequate to reach quasi-equilibrium. Therefore, the speed-up factor from this strategy is approximately a tenth of the total number of pulses. The accuracy of this strategy has been verified and validated for a wide range of conditions [25].

### 3.2 Correlated Dynamic Adaptive Chemistry and Transport (CO-DACT) Strategy

A fractional time-step Strang-splitting scheme [69] is employed to separate the convection-diffusion terms and the chemical source terms in the transport equations of neutral and ion species. The convection-diffusion terms are time-advanced using an explicit 4$^{th}$-order Runge-Kutta method. The chemical source terms are time-advanced implicitly using the classical variable coefficient stiff ODE solver (VODE) [70]. The subroutines for the plasma chemistry mechanism are generated from an open-source 0D plasma kinetics solver, ZDPlasKin [71]. Numerical diagnostics indicate that more than half of the computation time is spent on the implicit time-integration of chemical source terms. To tackle this issue, several acceleration techniques are proposed and employed with extensive accuracy verification [26].

Firstly, the point-implicit stiff ODE solver (ODEPIM) [26, 72-75], a semi-implicit solver, is employed to replace the purely implicit VODE solver. Providing similar accuracy, ODEPIM is approximately 60 times faster than VODE.

Since the computation time for chemistry is (linearly to cubically) proportional to the number of species, a dynamic adaptive chemistry (DAC) [76] technique is proposed. DAC generates a locally reduced kinetics mechanism for each spatial location and time step. Only the reaction rates of selected species and reactions are calculated, and the rest are frozen to reduce the calculation of chemistry. To reduce the large computational overhead for local mechanism reduction, a correlated version of DAC (CO-DAC) [26, 73-75, 77] is proposed to create time-space zones with similar thermo-chemical states, with only one time local reduction needed for each zone. CO-DAC is eminently suitable for the problems shown in Fig. 1, because of the uniformity of the bulk plasma and the quasi-periodic behaviors of the repetitive pulsed discharge, as shown in Figs. 3 and 4, respectively. CO-DAC further speeds up the chemistry calculation by a factor of approximately 3 with only negligible computational overhead.

Equipped with ODEPIM and CO-DAC, the dynamic evaluation of mixture-averaged transport coefficients (viscosity, thermal conductivity, and species diffusivity) becomes a large portion of the total computation time. Using the same correlation technique as in DAC but different zone grouping criteria [26, 73, 75, 78], the calculation of mixture-averaged transport coefficients is reduced by more than 800 times with negligible computational overhead.

### 3.3 Three-Stage Multi-Timescale Modeling Strategy

Plasma assisted ignition and combustion take place over a broad range of timescales. The ionization wave propagation, electrical breakdown, cathode sheath formation, and electron impact reactions have characteristic timescales at the sub-nanosecond level. Fast gas heating from the quenching of excited species and ion recombination has characteristic timescales on the order of microseconds. The cumulative effects of repetitive pulses, the convective and diffusive transport, and ignition and combustion, occur in the range from milliseconds to seconds.

In nanosecond pulsed plasma discharges, the sub-nanosecond scale physical processes occur only within the duration of each pulse. The microsecond level processes occur primarily during gaps between adjacent pulses. The millisecond to second scale processes occur generally after the accumulation of a significant number of pulses, and in burst mode, typically after all pulses are completed. To take advantage of this decoupling of timescales, a two-level adaptive time-step approach [24, 79] is proposed, and it is extended to a three-level version for computationally efficient simulation of ignition [57].

During each voltage pulse, all equations are solved simultaneously, and the time step size is set to O(0.1-1 ps). The exact time step size is dynamically determined based on the strength of E/N, the electron number density, and the Courant–Friedrichs–Lewy (CFL) condition for convection-diffusion terms.

In the duration between two consecutive pulses, both applied and gap voltages are nearly zero. As a result, the charge density quickly vanishes, and the electric field and its influence become negligible. To take advantage of this, the Poisson equation for electric potential and the electron energy equation are turned off, and the electric field and total charge are set to zero. This significantly reduces the stiffness of the system, such that the time step size can be increased to O(0.1-1 ns) without sacrificing accuracy or triggering numerical instability. The time step size is reset to O(0.1-1 ps) at the start of the following pulse, and this procedure is repeated until all pulses are finished.

After all pulses are completed, the time step of O(0.1-1 ns) is still too small to reach ignition in times on the order of 1 microsecond to 1 second within a reasonable computation time. To handle this issue, the transport equations of all charged species (electrons and ions) are turned off to further reduce the stiffness of the system, so that the time step, which is controlled by combustion chemistry, could be further increased to O(10-100 ns) [57] without sacrificing accuracy or triggering numerical instability.

## 4. General Theory

In this section, a general theory of plasma assisted ignition and combustion is developed, based on results from high fidelity modeling and simulation. The theory contains three components. The first component is a multi-timescale theory for nanosecond pulsed plasma discharges; this is the foundation for plasma enhancement on ignition and combustion. The second component is a theory of plasma assisted ignition, including one- and two- stage ignition, and the propagation of the ignition kernel. The last component is a model of plasma assisted flames, including both premixed and non-premixed flames.

### 4.1 Nanosecond Pulsed Plasma Discharges: a Multi-Timescale Theory

Previous studies [24, 28, 29, 38, 79, 80] on the nanosecond pulsed plasma discharge of air and fuel/oxidizer/diluent mixtures have provided very detailed information about the underlying physical processes. Here we bring these previous works together to develop a general theory.

Because of the multiscale nature of plasma, it is natural to classify its key physical processes based on their characteristic times. The characteristic timescale $\tau_k$ for the decay of the $k^{th}$ species can be estimated as $\tau_k = -\left[\frac{\partial}{\partial Y_k}\left(\frac{dY_k}{dt}\right)\right]^{-1} = \left(\frac{\partial Dec_k}{\partial Y_k}\right)^{-1}$, where the decay rate $Dec_k$ is defined as the rate sum of all reactions with the $k^{th}$ species as a reactant [81]. As an example, the temporal decay of electron, H, O, and OH is shown in Fig. 5 to indicate the different ranges of timescale involved. Since physical timescales depend on pressure (collision rates), the pressure range is specified as 60-150 Torr in this study. The pulsed discharges in this study have discharge durations of O(10 ns) and frequencies of 10-100 kHz. The gap between two adjacent pulses thus has a range of O(10-100 μs), and O(100) pulses could finish within a few ms.

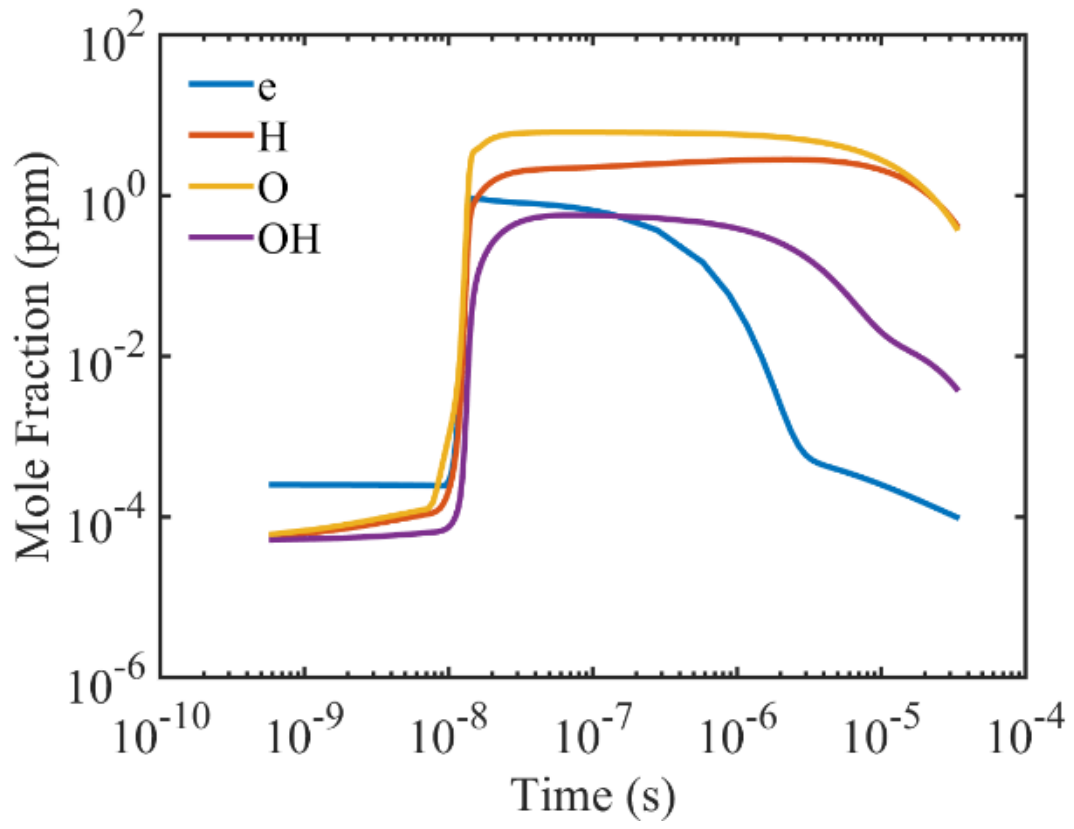

**Figure 5.** Time evolution of electron, H, O, and OH during 1st discharge pulse and 1st period (between beginning of 1st pulse and beginning of 2nd pulse) at center of domain. Simulation conducted in $C_2H_4/O_2/Ar$ mixture at 60 Torr initial pressure and 300 K initial temperature, at 30 kHz repetition rate. [28]

The multi-timescale theory for nanosecond pulsed plasma discharges is summarized in Table 1. The underlying physical processes are categorized into four stages with drastically different timescales. The first stage is the discharge pulse, which has timescales of O(1-10 ns). The second stage is the afterglow phase in the gap between two adjacent pulses, which has timescales of O(100 ns). The third phase is the rest of the gap between pulses, which has timescales of O(1-100 μs). The fourth stage is the completion of O(100) pulses, which has timescales of O(1 ms – 1 sec). Details of the physical processes in each stage will be discussed in the following sections.

**Table 1.** Multi-timescale theory for nanosecond pulsed plasma discharges

| Timescales (60-150 Torr) | Stage (10-100 kHz) | Key processes & mechanisms | Oxidizer & diluent | Fuel | Gas heating |
|---|---|---|---|---|---|
| O(1-10 ns) | Discharge pulse | Primary breakdown (60-75% of input energy coupling)<br>Fast ionization → uniform discharge<br>**Electron impact reactions**<br>(not $T_{gas}$ dependent) | Excitation (>50% of input energy)<br>Electron impact dissociation: $O_2 \rightarrow O + O(^1D)$ | H abstraction → H (re-activate LTC)<br>dissociation | Electron impact dissociation<br>Vibrational relaxation |
| | | | $O(^1D)$ + fuel:<br>• H abstraction → OH<br>• dissociation or pyrolysis | | |
| O(100 ns) | Gap between pulses<br>(afterglow) | e-ion recombination<br>**Quenching of excited species**<br>(weak T dependent) | Dissociation via quenching of excited species: $O_2 \rightarrow O + O(^1D)$ | dissociation | Fast gas heating<br>(0.5-2 K per pulse) |
| O(1-100 µs) | Gap between pulses | **Chain propagation & termination**<br>(sensitive to $T_{gas}$) | $O_2$ + fuel fragment / H → $HO_2$<br>$O/OH + HO_2 \rightarrow O_2 + OH/H_2O$<br>O/OH + fuel:<br>• H abstraction → $OH/H_2O$<br>• dissociation or pyrolysis<br>• Partial oxidation | | Exothermic chain termination |
| O(1 ms – 1 sec) | O(100) pulses<br>or<br>after all pulses | **Slow/weak process** (sensitive to $T_{gas}$)<br>Convection & diffusion | • $O + O_2 + M \leftrightarrow O_3 + M$<br>• dissociation or pyrolysis<br>• Partial oxidation<br>• $CO \rightarrow CO_2$ | | exothermic $O_3$/CO formation & partial oxidation |
| | | **Cumulative effects of multiple pulses** (weak $T_{gas}$ dependent) | Large pool of radicals & fuel fragments (enhance/re-activate LTC) | | Uniform 'hat-shaped' $T_{gas}$ profile |

### 4.1.1 Stage I – discharge pulse

During this stage, the discharge pulse deposits electrical energy into the gas mixture, which becomes the source of the non-equilibrium plasma. The energy deposition depends primarily on the ionization rate of the background gas. For most mixtures of fuel/oxidizer/diluent, the dominant background gases are oxidizer and diluent, because of the low concentration of fuel (typically a few percent level). Therefore, a change of fuel type has very limited influence on the energy deposition [38]. For a fixed peak voltage $V_{peak}$, the electrical energy deposition per pulse $Q_{pulse}$ is linearly proportional to the capacitor current, which is equal to the capacitance times the time rate of change in voltage $\left(\frac{V_{peak}}{t_{pulse}}\right)$ across the capacitor. In particular, based on a parametric study for $V_{peak}$ from 10 kV to 90 kV, Nagaraja and Yang [38] proposed that $Q_{pulse} = 0.14\left(\frac{V_{peak}}{t_{pulse}}\right) + 0.64$ for a pressure of 50 Torr and an initial gas temperature of 300 K. $Q_{pulse}$ is also proportional to pressure and thus particle number density, but it is only a weak function of pulsing frequency [24]. In contrast, the range of E/N depends strongly on pulsing frequency [80]. More precisely, as shown in Fig. 6, higher frequency has lower peak value for the primary spike of the E/N profile but higher peak values for the secondary spikes of the E/N profile. This is because higher pulsing frequency increases the residual electron density and serves as a uniform pre-ionization source, which lowers the breakdown voltage.

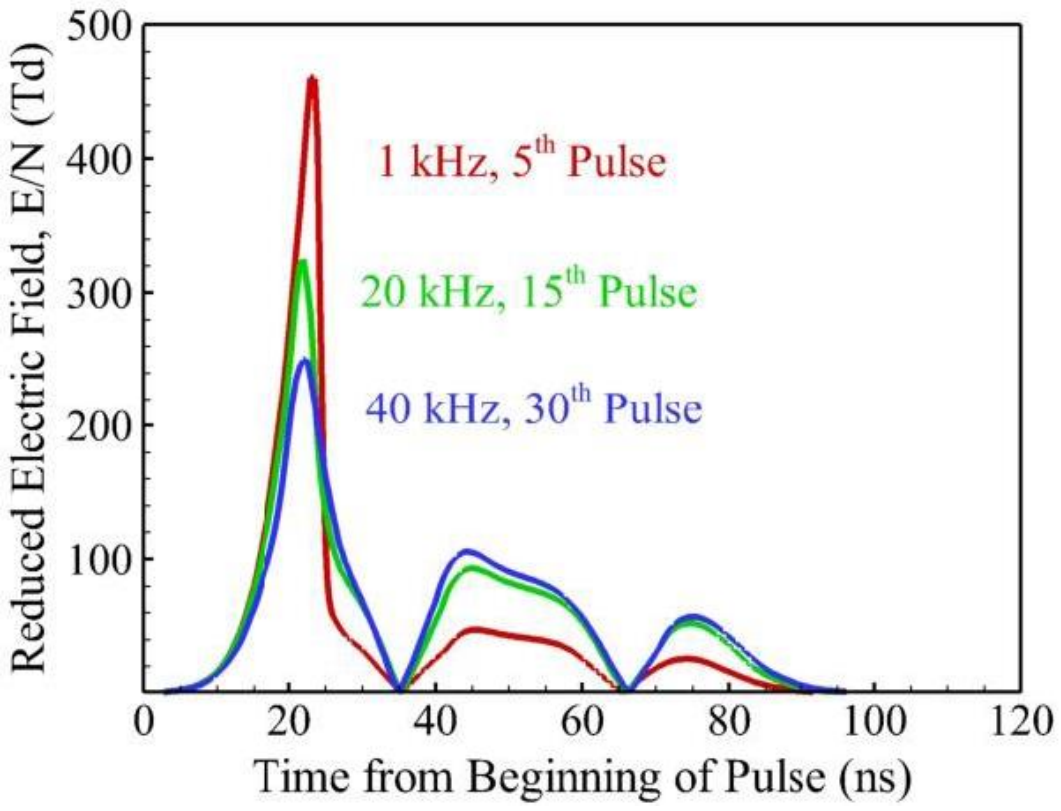

**Figure 6.** Temporal evolution of E/N during a nanosecond voltage pulse at different repetition rates after the discharge process has attained a periodic steady state. Simulation conducted in air at 60 Torr initial pressure and and 300 K initial temperature. [80]

Figure 7 shows that vibrational excitation dominates the input energy coupling when E/N is smaller than ~100 Td, while electronic excitation dominates when E/N is greater than ~100 Td [24]. For this reason, secondary spikes in the E/N profile represent pure vibrational energy coupling, and higher frequency pushes more input energy into vibrational excitation, which quickly relaxes and contributes part of the fast gas heating for bulk plasma [82].

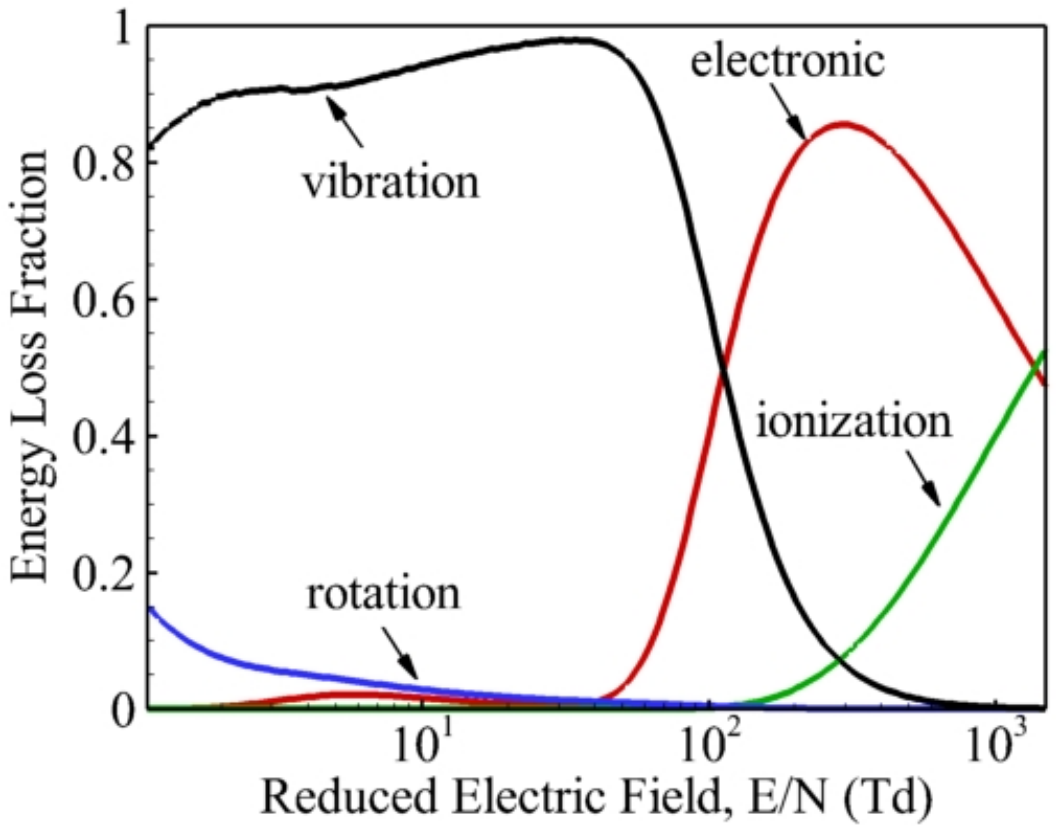

**Figure 7.** Fraction of electron energy lost in excitation of internal energy modes and ionization of $O_2$ and $N_2$ molecules in air as a function of E/N [24].

During the high voltage pulse, the fast ionization wave during the primary breakdown (60-75% of input energy coupling [24, 27, 28]) results in nearly uniform plasma discharge. The charge surge inside the sheath layers form a strong shield, which makes $V_{gap}$ and the electric field in the bulk plasma drop almost immediately after breakdown (Fig. 2). As shown in Fig. 3, the electric field inside the sheath layers continues to increase with further increase in applied voltage, which results in rapid ion Joule heating inside the sheath layers (30-40K within 80 ns) [24]. Under the high E/N inside the sheath layers, the electrons are accelerated significantly to fly out of the sheath layers into the bulk plasma with high energies of 10-100 eV [31].

The large number of high-energy electrons triggers ultra-fast electron impact reactions, which are the most important physical processes in this stage. Electron impact reactions are not dependent on gas temperature, and this is partially why plasma can trigger/sustain low temperature ignition/combustion. These reactions can be classified into two categories: electron impact excitation and electron impact dissociation. Electron impact excitation uses more than half of the input energy [31] and efficiently generates electronically excited $O_2$ and $N_2$/Ar species, including $O_2(a^1\Delta)$, $O_2(b^1\Sigma)$, $O_2(c^1\Sigma)$, $N_2(A^3\Sigma)$, $N_2(B^3\Pi)$, $N_2(C^3\Pi)$, $N_2(a'^1\Sigma)$, $Ar^*$. In most cases, the concentration of diluent ($N_2$/Ar) is higher than $O_2$, so the concentration of excited $N_2$/Ar is also higher than that of excited $O_2$. Electron impact dissociation reactions dissociate $O_2$ into O, $O(^1D)$, and $O(^1S)$, which contributes 40-70% of O generation and part of the fast gas heating in bulk plasma [24, 28]. $O+O(^1D)$ is the dominant product combination during the electron impact dissociation of $O_2$ [63]. Electron impact can also dissociate $N_2$ into N and $N(^2D)$ [58], but the rates of these dissociation reactions are significantly lower than those of $O_2$ because of the extremely strong triple bond in $N_2$. The electron impact on fuel molecules can trigger H-abstraction, dissociation/pyrolysis of fuel. The electron impact reaction dissociates $H_2$ fuel into two H radicals, which partially replaces the original chain initiation reactions in classical $H_2$ combustion chemical kinetics with high activation energies [38]. The electron impact reactions abstract H from $C_2H_4$ fuel molecules to produce fuel fragments of $C_2H_3$ and $C_2H_2$, and this contributes more than half of the H radical generation [28]. For fuels with typical LTC, electron impact reactions abstract H from fuel molecule RH (for example, fuel radical R is $C_7H_{15}$ for n-heptane, and is $CH_3OCH_2$ for DME) to produce R and H radicals, which can re-activate LTC for relatively higher initial gas temperatures above the typical LTC threshold temperature (such as 800 K at a reduced pressure of 76 Torr) [35, 57].

The fast reactions between $O(^1D)$ and fuel can result in further H-abstraction and dissociation/pyrolysis of fuel molecules, which are approximately 4 times faster than the quenching of $O(^1D)$ to O radical [28]. For $H_2$ fuel, a low temperature chain propagation reaction $O(^1D)+H_2 \rightarrow OH+H$ is introduced; it is more than two orders of magnitude faster than the classical chain propagation reaction $O+H_2 \rightarrow OH+H$ and dominates the OH radical production for the first few pulses [38]. For $C_2H_4$ fuel, the H-abstraction reaction $C_2H_4+O(^1D) \rightarrow C_2H_3+OH$ produces approximately 25% of OH radical generation; the dissociation reactions $C_2H_4+O(^1D) \rightarrow CH_2+CH_2O$ and $C_2H_4+O(^1D) \rightarrow CH_3+HCO$ produce 63% of $CH_2$, 17% of $CH_2O$ (a typical marker species for LTC), 60% of $CH_3$, and 20% of HCO [28]. For fuels with typical LTC, $O(^1D)$ abstracts H from fuel molecule RH to produce R and OH radicals, which contributes more than half of $O(^1D)$ consumption [29, 35].

#### 4.1.2 Stage II – gap between pulses (afterglow)

Quenching of $N_2(A^3\Sigma, B^3\Pi, C^3\Pi, a'^1\Sigma)$ or $Ar^*$ from Stage I further dissociates $O_2$ into O, $O(^1D)$, and $O(^1S)$, which contributes approximately 60% or 30% of O production, respectively [24, 28]. This indicates that either the production rate of excited $N_2$ is larger than that of excited

Ar, or the excited $N_2$ is more active than excited Ar. In addition, $O+O(^1D)$ is the dominant product during the dissociation of $O_2$ from quenching of excited diluent species. The quenching of $O(^1D)$ to O radical is approximately 4 times slower than its reactions with fuel molecules in Stage I [28]. Quenching of excited species also introduces several new $NO_x$ formation pathways: $N_2(A^3\Sigma)+O \rightarrow NO+N(^2D)$ and $N(^2D)+O_2 \rightarrow NO+O$ [58]. Although linearly proportional to the number of discharge pulses [53, 58], the amount of $NO_x$ introduced by these new pathways is much smaller than the decrease in $NO_x$ from the plasma assisted ultra-lean combustion [5]. The quenching of excited species can also result in fuel dissociation/pyrolysis. In particular, the quenching of $N_2(A^3\Sigma)$ can dissociate $H_2$ into 2H, which partially replaces the original chain initiation reactions in classical $H_2$ combustion chemical kinetics with high activation energies [38]. Quenching of the excited species is the major source for fast gas heating of 0.5-2 K per pulse inside the bulk plasma [27, 80, 82], and consumes 20-40% of the discharge energy [28, 82].

In this stage, in addition to quenching, there may be conversion of one excited species into another excited species. The major conversion pathways are $O(^1S) \rightarrow O(^1D)$ and $O_2(c^1\Sigma) \rightarrow O_2(b^1\Sigma) \rightarrow O_2(a^1\Delta)$ [28].

As shown in Fig. 5, the electron-ion recombination reactions have a timescale of O(100 ns) [28, 31], which is defined as the time for electron number density to reach its initial levels, thus occur primarily in this stage. Electron-ion recombination reactions contribute part of the fast gas heating of the bulk plasma, and some of them may trigger further decomposition: $e+O_2^+/N_2^+ \rightarrow 2O/2N$ [31]. Reactions in this stage have only weak gas temperature dependence, which is partially why plasma can trigger and sustain low temperature ignition and combustion.

#### 4.1.3 Stage III – gap between pulses

In the third stage, $O_2$ reacts with fuel fragments (such as HCO) or H radical to generate $HO_2$, one of the most important species for LTC. For $H_2$, the three-body reaction $H+O_2+M \rightarrow HO_2+M$ is the primary source of $HO_2$ [27]. For hydrocarbon fuels, this reaction is still primary source under high pressures, but $HCO+O_2 \rightarrow CO+HO_2$ is the primary source of both $HO_2$ and CO under low pressures (that is, ~90% of $HO_2$ and >80% of CO under 60 Torr) [28]. $HO_2$ is quasi-stable and reacts much slower than the other radicals at low gas temperatures, so its reactions occur primarily in this stage. Highly exothermic radical reactions with $HO_2$ ($O/OH+HO_2 \rightarrow O_2+OH/H_2O$) are much faster than those with fuel molecules, and dominate the OH radical production after approximately 0.1 ms (that is, after a few pulses) and contributes part of the gas heating in the bulk plasma [27, 38]. In addition, the reaction rates of radicals with $HO_2$ are proportional to pressure, but insensitive to equivalence ratio [25]. $HO_2$ also plays a role in $NO_x$ conversion and generates more OH radical: $NO+HO_2 \rightarrow NO_2+OH$ [58].

In this stage, temperature sensitive reactions between radical and fuel further contribute to the H-abstraction, dissociation/pyrolysis, partial oxidation, and thus part of the gas heating in the bulk plasma. For $C_2H_4$ fuel, reactions between O and $C_2H_4$ include both H-abstraction ($O+C_2H_4 \rightarrow H+CH_2CHO$) and dissociation ($O+C_2H_4 \rightarrow CH_3+HCO$), both of which are also partial oxidation and release heat to the bulk plasma. Under high pressure or low temperature, reaction with $C_2H_4$ fuel consumes most of the OH radical, and the product $CH_2CH_2OH$ is quasi-stable [28]. For fuels with typical LTC, radicals (O/H/OH) abstract H from fuel molecule RH to produce R and $OH/H_2/H_2O$ and release heat to the bulk plasma, contributing approximately 50% of the consumption of O radical, and most production and consumption of OH radical [29, 35].

Over the timescales of O(1 μs), the increase in gas temperature inside the sheath layers from the ion Joule heating in Stage I is rapidly dissipated, mainly to the cold electrode wall through heat conduction [24, 82]. This dissipation prevents the overheating of the sheath layers and the development of ionization instabilities.

### 4.1.4 Stage IV – finish of O(100) pulses

Under low pressures, the conversion between O and $O_3$ via $O+O_2+M \leftrightarrow O_3+M$ is a relatively slow process, and can equal or exceed the production rate of O only at timescales of O(ms) [24]. In addition, the decay of $O_2(a^1\Delta)$ and $O_2(b^1\Sigma)$ is even slower than the formation and decomposition of $O_3$, on the order of O(1 sec) and O(10 ms), respectively [28]. For combustion systems in which plasma and flame are spatially decoupled (that is, not *in situ*), the relatively long lifetimes of O, $O_3$, and $O_2(a^1\Delta)$ under low pressures makes them important carriers of the enhancement from plasma [31]. First, the plasma zone generates large amounts of O, $O_3$, and $O_2(a^1\Delta)$. Then, the flow transports these species to the preheat zone of the flame. Finally, the high gas temperatures in the preheat zone decompose $O_3$ to generate more O radical, which significantly shortens the auto-ignition delays in the preheat zone and stabilizes the flame under ultra-lean conditions.

For $C_2H_4$ fuel, the $CH_2CHO$ produced from Stage III can further react with $O_2$ to dissociate into CO, $CH_2O$, and OH, and this contributes approximately 40% of OH radical generation [28]. Under low gas temperatures and short timescales, more than half of $CO_2$ production comes from $CH_2+O_2 \rightarrow CO_2+(H+H,H_2)$ [18, 28], rather than from the highly exothermic chain propagation oxidation reaction $CO+OH \rightarrow CO_2+H$, which is too weak and slow to release large amounts of heat under such conditions [38].

Convection and diffusion processes also occur in this stage.

At the end of approximately 100 pulses with timescales of O(1 ms), a large pool of radicals and fuel fragments has been formed in the bulk plasma, accumulated from all the processes of Stages I-III [24], and a uniform ‘hat-shaped’ gas temperature profile has been developed. Since most of processes have weak gas temperature dependence or are even independent of gas temperature, plasma can trigger and sustain low temperature ignition and combustion under ultra-lean combustion conditions.

### 4.2 Plasma Assisted Ignition

Previous studies on plasma assisted ignition [25, 27, 35, 57, 58, 82] provide very detailed information about the underlying physical processes. For fuels with LTC, when the initial gas temperature is low enough, ignition takes place in two stages: first stage ignition to “cool flame,” and second stage ignition to hot flame [35]. For fuels without LTC, there is no first stage ignition or “cool flame,” and ignition is equivalent to the second stage in the two-stage ignition. After second stage ignition, the ignition kernel propagates to form a relatively stable flame. Thus, the physical processes of plasma assisted ignition can be naturally divided into three stages: first stage ignition, second stage ignition, and the propagation of ignition kernel. The general theory for plasma assisted ignition is summarized in Table 2. The details of the physical processes of each stage will be discussed in the following sections. The pressure range of the present study is specified as 60-150 Torr, and the pulsed discharges have durations of O(10 ns) and frequencies of 10-100 kHz.

**Table 2.** General theory for plasma assisted ignition

<table>
<tr><th>Stage</th><th>Fuel type</th><th>Stage initial T</th><th>Stage flame T</th><th>$\frac{\tau_{self}}{\tau_{plasma}}$</th><th>Key processes & mechanisms</th></tr>
<tr><td rowspan="2">1st stage ignition to “cool flame”</td><td rowspan="2">Fuels with LTC (e.g. DME, n-heptane)</td><td>> ~700-800 K</td><td rowspan="2">Linear function of input energy & initial T</td><td>$+\infty$</td><td rowspan="2">Plasma → e, $O(^1D)$, O, H, and OH → ultra-fast H-abstraction of fuel → exothermic LTC chain branching positive feedback cycle</td></tr>
<tr><td>< ~700-800 K</td><td>Exponential function of input energy</td></tr>
<tr><td rowspan="4">(2nd stage) ignition to hot flame</td><td>Fuels with LTC</td><td>1st stage (“cool flame”) T</td><td rowspan="3">~ adiabatic flame T</td><td>Exponential to “cool flame” T & input energy after 1st stage</td><td rowspan="4">Fast gas heating<br>(0.5-2 K per pulse)<br><br>Plasma/LTC → radicals → exothermic T-sensitive chain branching positive feedback cycle:<br>• $HO_2 \rightarrow H_2O_2 \rightarrow 2OH$<br>• Fuel dissociation or pyrolysis<br>• Fuel oxidation</td></tr>
<tr><td rowspan="3">Fuels without LTC</td><td>> self-auto-ignition T</td><td>Exponential function of input energy</td></tr>
<tr><td>Between plasma & self- auto-ignition T</td><td>$+\infty$</td></tr>
<tr><td>< plasma auto-ignition T</td><td>< plasma auto-ignition T</td><td>No ignition</td></tr>
<tr><td>Ignition kernel propagation</td><td>All fuel types</td><td>Plasma → lower auto-ignition T</td><td>~ flame T uniformly</td><td>$S_{ign} \gg S_L$: diffusion not dominant</td><td>Plasma → local radical generation & preheating<br><br>(Smaller MIE)</td></tr>
</table>

### 4.2.1 First Stage Ignition to “Cool Flame”

Two-stage ignition and “cool flame” are only present in fuels with LTC, such as DME and n-heptane. First stage ignition to “cool flame” occurs through a self-sustaining exothermic LTC chain branching cycle (Fig. 8) [35]. The fuel molecule can be expressed as RH, where R is its corresponding fuel radical (R is $CH_3OCH_2$ for DME, and $C_7H_{15}$ for n-heptane). First, H abstract reaction of radicals (such as O, H, OH) from the fuel molecule RH generates fuel radical R. Then, R reacts with $O_2$ to form $RO_2$. $RO_2$ then isomerizes to hydroperoxy alkyl radical QOOH, where Q is the corresponding alkyl radical (Q is $CH_2OCH_2$ for DME, and $C_7H_{14}$ for n-heptane). After that, QOOH adds another $O_2$ to form $O_2$QOOH. Finally, $O_2$QOOH isomerizes and decomposes to generate two OH radicals, which complete the chain branching cycle.

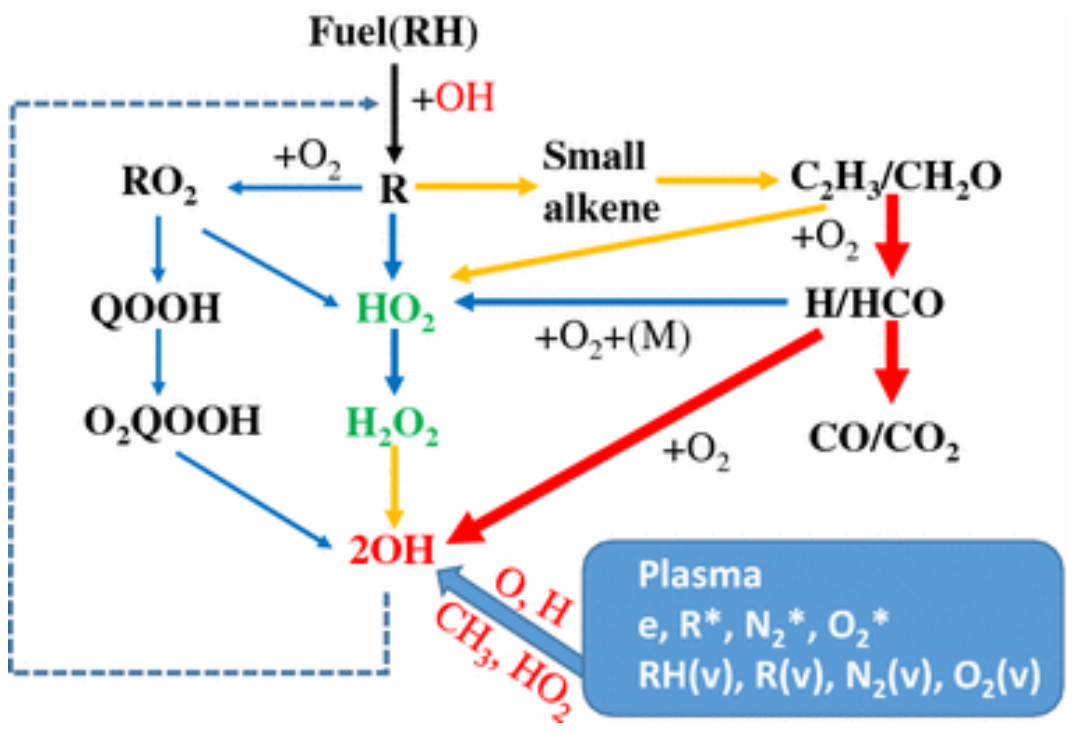

**Figure 8.** Schematic of the key reaction pathways for plasma assisted fuel oxidation at different gas temperatures (blue arrow below 700 K; yellow arrow 700–1050 K; red arrow above 1050 K). Note that critical gas temperature conditions are pressure dependent. e: energetic electrons, *: electronically excited molecules; v: vibrationally excited molecules. [35]

For initial gas temperatures below ~700-800 K and 150 discharge pulses (finishing at 5 ms), as shown in Fig. 9, the first stage ignition to “cool flame” can occur without plasma. Therefore, the plasma enhancement has a catalytic effect and depends only weakly on the number of pulses [58]. Plasma initiates a small amount of seed radicals, such as O, H and OH, at the beginning, and this accelerates the H abstraction reactions from fuel molecules and triggers the self-accelerating LTC feedback loop. This significantly reduces (approximately10 times) the induction time of the cycle. In this case (for a small number of pulses), the plasma enhancement of LTC is nearly independent of equivalence ratio. It is more pronounced at lower gas temperatures (550–650 K) because of the difficulty of initial radical production at low gas temperature conditions, where the LTC is stronger [58].

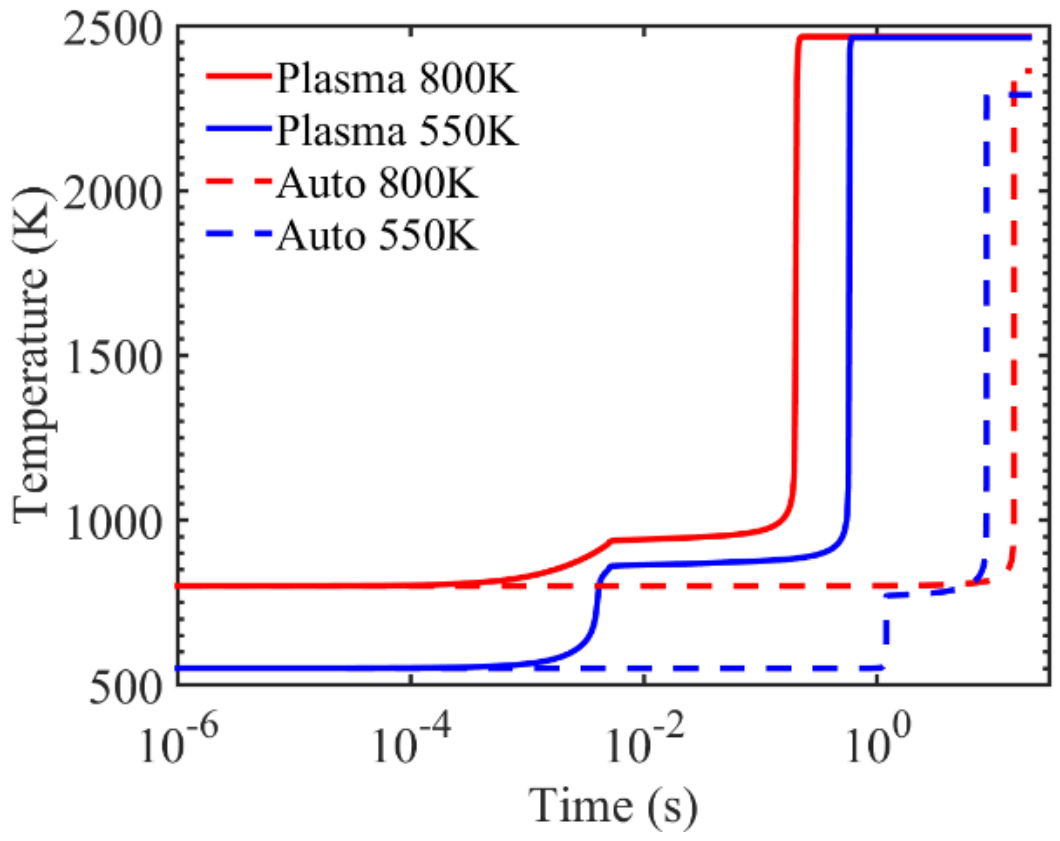

**Figure 9.** Comparison of gas temperatures for plasma assisted stimulation (30 kHz repetition rate) and auto-ignition of DME/$O_2$/Ar mixture at 76 Torr pressure [57].

For initial gas temperatures above ~700-800 K, as shown in Fig. 9, there is no first stage ignition or “cool flame” without plasma. This is because direct decomposition of fuel radical R into small alkenes (the yellow and red pathways in Fig. 8) is significantly accelerated at higher initial gas temperatures, and the LTC chain branching cycle, which has weak gas temperature dependence, is negligible. However, plasma discharges introduce ultra-fast H-abstraction reactions e/O($^1$D)+RH→R+(H+e)/OH to produce a large amount of R orders of magnitude faster and higher than regular H-abstractions via radicals [57] before ignition. Due to the limited decomposition rate of R, a significant amount of R goes to the LTC cycle again, and the first stage ignition to “cool flame” reappears, as shown in Fig. 9. The strength of this re-activated

LTC cycle is linearly proportional to the accumulative input energy, as is the "cool flame" temperature [57].

For all gas temperature ranges, when the number of pulses is large (approximately 50 pulses or more), plasma enhancement is more than just a catalytic effect, and the relationship between "cool flame" temperature and accumulative input energy is linear even when the initial gas temperature is below ~700-800 K. In this case, the normalized first stage ignition rate $\frac{\tau_{self}}{\tau_{plasma}}$ (the reduction factor of the first stage ignition delay) is an exponential function of accumulative input energy [57]. The reason for this is that the reactivity to generate radicals and excited species is an exponential function of electron temperature at low gas temperatures, which is linearly proportional to input electrical energy. In this case, the reduction of the first stage ignition delay is much larger than in the pure catalytic case; for example, 150 pulses reduce ignition delay by approximately 250 times for a DME/$O_2$/Ar mixture at 76 Torr [57].

### 4.2.2 Second Stage Ignition to Hot Flame

The concept of the second stage ignition is relevant to the two-stage ignition of fuels with LTC; the ignition of fuels without LTC can be considered equivalent to the second stage ignition. This stage can be further divided into two steps. In the first step, the gas-heating sources increase the gas temperature to a threshold value (auto-ignition temperature). Below this threshold gas temperature, ignition will not occur because of the balance of quenching and radical production. In the second step, above the threshold gas temperature, chain branching and fuel oxidation pathways are triggered and accelerated until the process becomes self-sustaining, and the gas temperature can continue to rise until the final ignition occurs. For this reason, during this second step, the effect of plasma is negligible [27]. The threshold value exists because typical chain branching reactions are endothermic and highly gas temperature sensitive. A small increase in gas temperature near ignition significantly increases the chain branching reaction rates to produce radicals, which then accelerates the heat release from fuel oxidation. Consequently, the radical profiles are much steeper than the gas temperature profile. The increase in gas temperature completes the cycle by further increasing the rates of chain branching reactions. This non-linear positive feedback cycle causes an exponential increase in both gas temperature and the concentration of radicals leading to ignition.

The presence of plasma can reduce the threshold auto-ignition temperature and enhance or shorten the first step but not the second step. For example, under pressures of 80-100 Torr, plasma reduces the auto-ignition temperature of a lean $H_2$/air mixture (equivalence ratios of 0.06-0.12) from ~900 K to ~700 K, as shown in Fig. 10 [27]. As described in previous sections, fast gas heating of 0.5-2 K per pulse takes place in the bulk plasma, primarily from the quenching of excited species. In addition, the large pool of radicals and excited species generated by the plasma or the LTC chain branching cycle pre-activates the high-temperature chain branching and oxidation pathways (such as $2HO_2 \rightarrow H_2O_2 + O_2$ and $H_2O_2 \rightarrow 2OH$), which also release a large amount of heat. Therefore, for a given type of nanosecond pulsed discharge, there is a threshold number of pulses, or equivalently a threshold accumulative input energy, to trigger ignition. Plasma is thus seen to have a larger enhancing effect in the first stage [57, 58], where it can have catalyst effects, while the second stage depends on purely threshold-like behavior. For two-stage ignition of fuels with LTC, if the number of pulses after the first stage is small (20-30 pulses), then the radical pool directly introduced by plasma will be much smaller than that introduced by the LTC cycle of the “cool flame.” In that case, a plasma burst after the first stage ignition provides predominantly thermal enhancement, which only reduces the second stage ignition delay by approximately 30% [58].

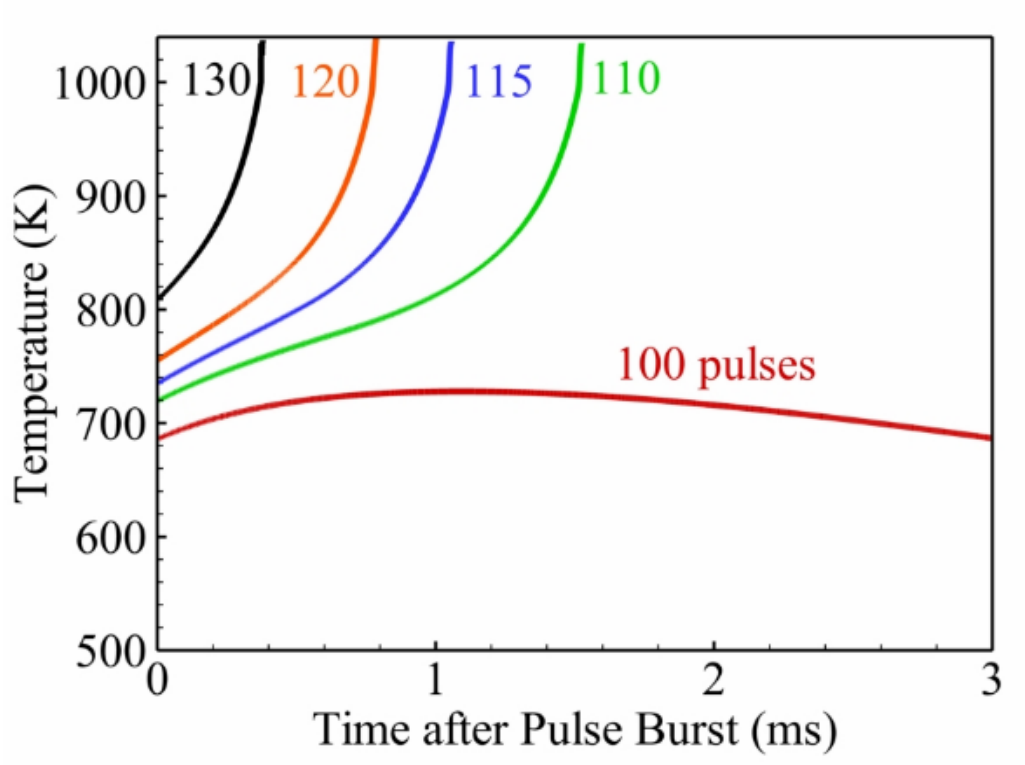

**Figure 10.** Temporal evolution of gas temperature at center of discharge volume. Initial pressure and gas temperature are 80 Torr and 473 K, respectively ($H_2$/air mixture, CPT pulser, 40 kHz). [27]

As shown in Fig. 10, the ignition delay increases steeply as the number of pulses is reduced in the plane-to-plane discharge system. The normalized ignition rate $\frac{\tau_{self}}{\tau_{plasma}}$ (reduction in ignition delay) is an exponential function of accumulative input energy [27], because reactivity is an exponential function of electron temperature at low gas temperatures, and an exponential function of gas temperature at high gas temperatures, both of which are linearly proportional to input electrical energy. For fuels with LTC, this relation is equivalent to an exponential function of both "cool flame" temperature and the accumulative input energy after the first stage [57], because the "cool flame" temperature is linearly proportional to the accumulative input energy before first stage ignition. When the number of pulses is small (within ~10 pulses), the discharge pulses applied at the beginning have negligible impact on the "cool flame" temperature, and thus negligible impact on the second stage ignition delay [58]. In contrast, when the number of pulses is large (approximately 50 pulses or more), the LTC chain branching cycle is either significantly enhanced or re-activated, which generates a much larger radical pool and heat release than the un-enhanced or un-activated cycle and significantly reduces the second stage ignition delay. Taking a DME/$O_2$/Ar mixture under 76 Torr and 150 discharge pulses as an example [57], for initial gas temperatures below ~700-800 K, the LTC chain branching cycle is only enhanced, so the ignition time reduction factor is only ~15. For initial gas temperatures above ~700-800 K, the LTC chain branching cycle is re-activated, so the reduction factor could be as large as ~75.

### 4.2.3 Ignition Kernel Propagation

Auto-ignition first occurs at the center of the discharge gap, where the gas temperature first reaches the reduced threshold value [27]. Nevertheless, assisted by plasma discharges, the gas temperature profile is close to uniform distribution [24, 28], such that other locations also reach the reduced threshold gas temperature rapidly (indeed, almost simultaneously). Auto-ignition occurs independently at different locations because the local preheated mixture is doped with radicals from the plasma discharge burst [82], such that heat transport does not play a significant role [27]. This is different from hot-spot thermal igniters, which require diffusion of both radicals and heat [82]. As a result, the ignition kernel expands rapidly to the entire volume, except near the quasi-isothermal walls, where the rapid heat conduction losses from high gas temperature gradients keep the gas temperature low [24, 28]. Therefore, the plasma assisted ignition kernel propagation speed is much faster than laminar flame speed (approximately 15 times faster for lean $H_2$/air mixture at 80 Torr), and plasma can ignite the entire volume of premixed mixture much faster than thermal ignition but with smaller minimum ignition energy (MIE) [82]. This is due to the more efficient excited species/radical production from plasma discharges, as described in Section 4.1.

## 4.3 Plasma Assisted Flames

Previous studies on plasma assisted flames covered the two basic flame types: premixed [30] and non-premixed (diffusion) flames [31]. All other (practical) flames are combinations of these two basic flame types, so the plasma enhancement on the other flame types can be inferred. These studies provide very detailed information on the underlying physical processes; here we organize them to develop general theories for both basic flame types.

### 4.3.1 Premixed Flame

In this study, due to the limitations of the 1D model, we only consider the laminar flat premixed flame. The flame is enhanced by *in situ* plasma discharges, as shown in Fig. 1(b). The premixed flame contains several zones: the pre-heat, reaction, and post-flame zones. In the laminar premixed flame, the reaction zone is very thin, so the volumetric plasma enhancement is not significant in the reaction zone. Near the burner, due to fast heat conduction to the wall, the gas temperature is lower than ~500 K, and several quasi-stable species could be formed and last. Based on these analyses, therefore, we organize the general theory for plasma assisted premixed flame into three zones, neglecting the reaction zone: the near-the-burner, pre-heat, and post-flame zones. The theory is summarized in Table 3.

**Table 3.** General theory for plasma assisted premixed flame

<table>
<tr><td colspan="2">Zone</td><td>Near the burner</td><td colspan="2">Pre-heat zone</td><td>Post-flame zone</td></tr>
<tr><td colspan="2">T range</td><td>< ~500 K</td><td>~500-700 K</td><td>~700 K – flame T</td><td>Flame T</td></tr>
<tr><td colspan="2">$\rho$ & N</td><td colspan="3">high</td><td>low</td></tr>
<tr><td colspan="2">E/N range</td><td colspan="3">~100-700 Td</td><td>~700-1500 Td</td></tr>
<tr><td colspan="2">Dominant mode</td><td colspan="3">Electronic</td><td>Ionization</td></tr>
<tr><td colspan="2">Input energy</td><td colspan="3">~10%</td><td>~90%</td></tr>
<tr><td rowspan="5">Non-equilibrium plasma chemistry:<br>• Spatially uniform<br>• Gradient ↑</td><td>$O_3$</td><td>$O_2+O+M \rightarrow O_3+M$</td><td colspan="2">$O_3(+M) \rightarrow O_2+O(+M)$</td><td rowspan="4">electron impact ionization does NOT contribute radical production</td></tr>
<tr><td>$HO_2$</td><td colspan="2">$H+O_2+M \rightarrow HO_2+M$</td><td>$O/OH+HO_2 \rightarrow O_2+OH/H_2O$</td></tr>
<tr><td>O</td><td colspan="3">Increase by a factor of ~6</td></tr>
<tr><td>H</td><td colspan="3">Increase by a factor of ~4</td></tr>
<tr><td>OH</td><td colspan="3">Increase by ~100-500%</td><td>Increase by ~40%</td></tr>
<tr><td colspan="2">Direct gas heating</td><td colspan="4">A significant portion of plasma enhancement</td></tr>
<tr><td colspan="2">Enhancement for T</td><td colspan="4">Increase by ~20%; gradient ↑</td></tr>
</table>

By definition, E/N varies inversely with N and $\rho$ during each pulse. Therefore, due to intense gas expansion from heat release, high E/N (~700–1500 Td) regions are found downstream of the flame (in the high gas temperature post-flame zone). Inside the cathode sheath (near the boundary of the post-flame zone), E/N attains its highest value of approximately 1500 Td for a short duration. In this high E/N range, as shown in Fig. 7, a significant fraction of the input energy goes to ionization, and thus less electron energy can go to electronic mode to generate radicals via electron impact excitation and dissociation. In contrast, lower E/N values (100–700 Td) in the lower gas temperature pre-heat zone and near the burner push more electron energy into electronic mode, and this promotes more electron impact excitation and dissociation to generate radicals and excited species in an efficient and spatially uniform manner. As a direct comparison, plasma increases OH concentrations by ~100-500% in the pre-heat zone, but only by ~40% in the post-flame zone. The largest increase of ~500% occurs near the burner surface. Increased OH concentrations enhance the heat release from the primary chain propagating oxidation reactions ($OH+H_2/CO \rightarrow H_2O/CO_2+H$). The resulting increase in gas temperature accelerates the conventional fuel oxidation chain branching processes, creating a positive feedback loop for rapid production of radicals. With plasma discharge, gas temperatures increase by approximately 20% in both the pre-heat and post-flame zones. Effects of both chemistry and direct gas heating are responsible for a significant portion of the plasma enhancement of the flame. In fact, non-equilibrium plasma generates considerably higher amounts of radicals than the same rate of indiscriminate gas heating. The plasma chemistry enhancement of the flame is more pronounced at fuel-lean conditions.

In the pre-heat zone, plasma discharges significantly increase O and H radicals, with the peak values increasing by factors of six and four, respectively. In addition, gradients of radical concentrations and gas temperature are also substantially higher. Both species and gas temperature profiles shift upstream toward the burner at low plasma power (approximately 0.2 cm), suggesting that plasma accelerates auto-ignition in the pre-heat zone. Closer to the burner, the low gas temperature enhances the formation of $O_3$ and $HO_2$ via three-body reactions from O and H radicals, respectively, although this process is slower than the reactions between O and fuel. $O_3$ travels a short distance downstream until its decomposition when the gas temperature rises to ~500 K. $HO_2$ persists for a longer distance, until the gas temperature increases to above ~700 K.

In this configuration, only ~10% of the total input energy is coupled in the pre-heat and reaction zones, while ~90% of the total input energy is coupled in the post-flame zone and does not enhance the pre-heat zone. To increase the energy coupled in the pre-heat zone, it has been proposed that the high-voltage electrode should be placed closer to the flame, which will require that the electrode material tolerate higher temperatures.

### 4.3.2 Non-Premixed (Diffusion) Flame

In this study, a counter-flow diffusion flame is employed at pressures ranging from 60 to 300 Torr, with the oxidizer stream enhanced by plasma discharges. As shown in Fig. 1(c), this configuration can be divided into four zones: the plasma discharge zone inside the oxidizer tube, the region near the oxidizer tube exit, the radical production zone between the oxidizer tube exit and the flame, and the flame zone. Based on this analysis, a general theory for a plasma assisted non-premixed (diffusion) flame is developed, and summarized in Table 4.

**Table 4.** General theory for plasma assisted non-premixed (diffusion) flame

<table>
<tr><th>Zone</th><th>T</th><th>Chemistry</th></tr>
<tr><td>Discharge</td><td>Initial T</td><td>Afterglow: $O_2$+O+M→$O_3$+M</td></tr>
<tr><td>Oxidizer tube exit</td><td>Plasma → preheated T</td><td>Only long lifetime species survive: O, $O_3$, $O_2(a^1\Delta)$</td></tr>
<tr><td>Radical production zone: shift upstream</td><td>Preheated T – flame T</td><td>$O_3$(+M)→$O_2$+O(+M) rapidly; $O_2(a^1\Delta)$→$O_2$ rapidly; only O survives</td></tr>
<tr><td rowspan="2">Flame zone: shift upstream towards the oxidizer tube exit</td><td>Flame T ↑ (dominant)</td><td>O enhances chain branching: O+$H_2$→OH+H; O+$H_2O$→2OH ⇒ OH ↑ ~10 times (larger at low p)</td></tr>
<tr><td colspan="2">⇒ rapid fuel consumption: breakup, chain branching, oxidation (OH+$H_2$/CO→$H_2O$/$CO_2$+H)</td></tr>
</table>

In this configuration, the plasma discharges are not *in situ* with respect to the flame, so only species with long lifetimes of O(1 ms), mainly including O, $O_3$, and $O_2(a^1\Delta)$, can reach the oxidizer tube exit. Among them, $O_3$ is primarily generated in the afterglow by three-body recombination $O_2+O+M \rightarrow O_3+M$. At higher pressures, the recombination rates of plasma generated radicals increase such that fewer radicals could survive to reach the flame zone from the discharge zone. On the other hand, plasma discharges also pre-heat the oxidizer flow to higher gas temperature.

In the radical production zone, due to the higher gas temperature, $O_3$ rapidly decomposes to $O_2$ and O radical, and $O_2(a^1\Delta)$ rapidly quenches to $O_2$. As a result, only O radical can reach the flame to provide enhancement from plasma.

Due to the plasma generated O radical, the OH radical increases by approximately 10 times in the flame zone, primarily through chain branching reactions of $O+H_2 \rightarrow OH+H$ and $O+H_2O \rightarrow 2OH$. This increase in OH radical, together with the higher oxidizer flow temperature preheated by the plasma, results in rapid fuel consumption in the flame zone, including breakup, chain branching, and oxidation ($OH+H_2/CO \rightarrow H_2O/CO_2+H$). In addition, the peak flame temperature also rises noticeably with plasma activation.

If the flame location is identified by maximum temperature location, then plasma shifts both the radical production zone and flame location toward the oxidizer tube exit. Plasma also significantly increases the extinction strain rates by more than two times, to sustain stable lean combustion.

## 5. Conclusions

A self-consistent theoretical framework for plasma assisted ignition and combustion is reviewed. Three physical configurations are considered to cover plasma assisted ignition, and plasma assisted premixed and non-premixed flames. A two-temperature model is employed for non-equilibrium plasma. It is seen that modeling must be at least one-dimensional to capture cathode sheath formation, and to calculate E/N accurately. The plasma-combustion chemistry mechanisms for a wide range of fuel types are reviewed to illustrate how to develop such mechanisms from existing air plasma chemistry and fuel combustion chemistry.

Novel modeling strategies for plasma assisted ignition and combustion are reviewed and summarized. The "frozen electric field" strategy takes advantage of the quasi-periodic behaviors of the electrical characteristics to avoid re-calculation during each pulse. To deal with the intense calculation of large and stiff chemistry and the evaluation of mixture-averaged transport properties, a point-implicit stiff ODE solver (ODEPIM) and correlated dynamic adaptive chemistry and transport (CO-DACT) strategy are employed. CO-DACT generates locally reduced kinetics mechanisms for each spatial location and time step. Only the reaction rates of selected species and reactions are calculated, while the rest are frozen to reduce the chemistry calculation. A three-stage multi-timescale modeling strategy takes advantage of the separation of timescales in nanosecond pulsed plasma discharges to dynamically adjust the time step size for simulations.

A general theory of plasma assisted ignition and combustion is proposed. The theory has three major components: a multi-timescale theory for nanosecond pulsed plasma discharges, a theory for plasma assisted ignition, and theories for plasma assisted premixed and non-premixed flames.

A multi-timescale theory for nanosecond pulsed plasma discharge sets the foundation of plasma enhancement for ignition and combustion; it can be divided into four stages based on the timescales of the underlying physical processes. Stage I is the discharge pulse, which has timescales of O(1-10 ns). In this stage, most input energy is coupled into electron impact excitation and dissociation reactions, to generate a large amount of excited species and radicals under low gas temperatures. Stage II is the afterglow phase during the gap between two adjacent pulses, and has timescales of O(100 ns). Quenching of excited species is the primary physical process in this stage, and not only further dissociates both $O_2$ and fuel molecules to generate radicals under low gas temperatures, but also provides fast gas heating. Stage III is the remainder of the gap between two adjacent pulses, with timescales of O(1-100 μs). The radicals generated during Stages I and II significantly enhance the exothermic chain propagation and termination reactions in this stage. Stage IV takes place after a large number of pulses, with timescales of O(1 ms – 1 sec). This stage includes weak and slow reactions, convective and diffusion transport, and the cumulative effects of multiple pulses: preheated gas temperatures and a large pool of radicals and fuel fragments.

Plasma assisted ignition can be divided into three stages: the first stage ignition to "cool flame" (only for fuels with LTC), the second stage ignition to hot flame, and the ignition kernel propagation. In the first stage, plasma generates a large amount of e, $O(^1D)$, O, H, and OH, enhancing the H-abstraction of fuel molecules, which in turn either significantly enhances (below ~700-800 K) or re-activates (above ~700-800 K) the exothermic LTC chain branching positive feedback cycle. In the second stage, the radicals from the first stage trigger and enhance the exothermic high gas temperature chain branching cycle, which leads to ignition of hot flames. The acceleration of ignition is a function of input energy. The ignition kernel propagation is close to auto-ignition triggered by radicals and pre-heating at different locations; it is much faster than laminar flame speed and requires lower input energy than purely thermal ignition.

For plasma assisted premixed and non-premixed flame, plasma significantly enhances the radical generation and gas heating in the pre-heat zone, which triggers ultra-fast auto-ignition and flashback.

## Acknowledgements

This work was supported by the William R.T. Oakes Endowment of the Georgia Institute of Technology.